\documentclass[aip,jcp,reprint]{revtex4-1}
\usepackage{graphicx}

\draft 

\begin{document}


\title{Capillary Condensation in Cylindrical Pores: Monte Carlo Study of the Interplay of Surface and
Finite Size Effects} 



\author{A. Winkler}
\author{D. Wilms}
\author{P. Virnau}
\author{K. Binder$^{\text{*}}$}
\affiliation{$^{\text{*}}$ Institut f\"{u}r Physik, Johannes Gutenberg-Universit\"{a}t, D-55099 Mainz, Staudinger Weg 7, Germany}


\date{\today}

\pacs{64.75Jk, 64.60.an, 05.70Fh, 02.70Tt}

\begin{abstract}
When a fluid that undergoes a vapor to liquid transition in the
bulk is confined to a long cylindrical pore, the phase transition
is shifted (mostly due to surface effects at the walls of the
pore) and rounded (due to finite size effects). The nature of the
phase coexistence at the transition depends on the length of the
pore: For very long pores the system is
axially homogeneous at low temperatures. At the chemical potential where the
transition takes place fluctuations occur between vapor-like and
liquid-like states of the cylinder as a whole. At somewhat
higher temperatures (but still far below bulk criticality) the
system at phase coexistence is in an axially inhomogeneous
multi-domain state, where long cylindrical liquid-like and
vapor-like domains alternate. Using Monte Carlo simulations for
the Ising/lattice gas model and the Asakura-Oosawa model of
colloid-polymer mixtures the transition between these two
different scenarios is characterized. It is shown that the density
distribution changes gradually from a double-peak structure to a
triple-peak shape, and the correlation length in axial direction
(measuring the equilibrium domain length) becomes much smaller
than the cylinder length. The (rounded) transition to the
disordered phase of the fluid occurs when the axial correlation
length has decreased to a value comparable to the cylinder
diameter. It is also suggested that adsorption hysteresis vanishes
when the transition from the simple domain state to the multi-domain
state of the cylindrical pore occurs. We predict that the difference between
the pore critical temperature and the hysteresis critical temperature should
increase logarithmically with the length of the pore.
\end{abstract}

\pacs{}

\maketitle 

\section{Introduction}

The properties of both pure fluids and fluid mixtures confined to
nanoporous and microporous materials \cite{1,2,3} have found a lot of
interest recently, both from the point of view of various
applications \cite{4,5,6,7,8,9,10,11}, and also because phase
transitions in confined geometry are a problem of fundamental
importance in statistical thermodynamics
\cite{1,2,3,12,13,14,15,16}. Applications range from the technique
to extract oil and gas from porous natural rocks, the use of
artificial mesoporous materials such as various zeolithes as
catalysts, ``molecular sieves'' to separate fluids in fluid
mixtures, and various microfluidic and nanofluidic devices
\cite{1,2,3,4,5,6,7,8,9,10,11}. While in some cases (e.g. vycor
glass \cite{17,18}) the random irregularity of the porous network
structure is expected to lead to important physical effects
\cite{19}, one can also study the idealized case of isolated long
straight pores experimentally, both for pore widths on the scale of nanometers
(e.g. filling fluids into carbon nanotubes \cite{20,21}) and for pore widths on
the scale of up to $150\mu m$ (producing arrays of such pores in
silicon wafers \cite{22}, e.g.~for the purpose of characterization
of DNA put into such pores \cite{23}, etc).

Since a long time it is known that the vapor to liquid transition
in pores is typically shifted relative to the condition where it
occurs in the bulk: for lyophilic pore walls the condensation
already occurs at a chemical potential where the vapor in the bulk
is still undersaturated (``capillary condensation'')
\cite{1,2,3,24,25,26,27,28,29,30}, but for lyophobic pore walls the opposite effect is also possible (``capillary evaporation'')
\cite{31,32,33,34,35,36}. To characterize these phenomena
quantitatively, however, one needs to understand the extent to
which wetting (or drying, respectively) phenomena
\cite{37,38,39,40,41,42,43,44} exist in this restricted
cylindrical geometry (obviously, infinitely thick wetting or
drying layers do not exist in narrow cylinders). An experimentally
important effect, that has also found a lot of theoretical
attention, is the ``adsorption hysteresis'' that obscures the true
equilibrium behavior of capillary condensation in pores, at least
over some range of parameters
\cite{4,28,45,46,47,48,49,50,51,52,53,54,55,56,57,58,59,60}.
Another question concerns the understanding of critical phenomena
when one reaches conditions where the density difference between
the vapor-like and liquid-like ``phases'' in the pore vanishes
\cite{1,2,3}. Here, one encounters a fundamental problem of
statistical mechanics, since the correlation length of density
fluctuations can show unlimited growth only along one direction
(the pore axis), but one does not at all expect any phase transition
for quasi-one-dimensional systems with short-range forces
\cite{61,62,63,64}. Nevertheless, a lot of phase diagrams and
critical points for various fluids confined in nanoscopic pores
have been quoted in the experimental literature
\cite{1,2,3,51,51,52,53,54,55,56,65} and in the theoretical work
\cite{1,2,3,28,57,58,66,67,68}. This fluctuation-induced
destruction of the phase transition is also not seen in
theoretical work based on density-functional theory \cite{28,30}
(or related mean-field theories), and cannot be seen in computer
simulations either, if one chooses pore lengths not much larger
than the pore diameter, as is done in many cases
\cite{1,2,3,68,69}, or if one constrains fluctuations by other
methods \cite{57,58}.

In the present work, we wish to contribute to the theoretical
understanding of these problems, presenting computer simulations
of two models, the Ising/lattice gas model confined in cylindrical
geometry (as well as its two-dimensional analog, Ising strips of
finite width), and the Asakura-Oosawa (AO) model for
colloid-polymer mixtures \cite{70}, confined in cylinders with
hard (infinitely repulsive) walls. A distinctive feature of our
work is that we pay detailed attention to the dependence of
various physical properties on the length $L$ of the cylinder,
confining attention to the (physically relevant) case $L \gg D$
throughout.

Sec.~2 presents a selection of our numerical results for the
Ising lattice, while Sec.~3 provides corresponding Monte Carlo data
for the off-lattice AO model, and discusses the
generic features of both models, interprets them in terms of
phenomenological theoretical considerations, and draws some
conclusions on pertinent experiments. Sec.~4 contains a brief
summary of our work.

\section{The order parameter distribution function $P_{L,D}(M)$ of
quasi-one-dimensional lattice gas models}
\subsection{Ising strips in the $L \times D$ geometry for $L \gg
D$}\label{Theory}

Ising (lattice gas) models in quasi-one-dimensional geometry have
already been considered extensively in the literature (e.g.
\cite{62,63,71,72,73,74,75,76,77,78,79,80,81,82,83,84,85,86,87,88})
but here we focus attention to an aspect which (to our knowledge)
has not been studied before, namely the relation between the
correlation length $\xi$ in the long direction of the strip and
the distribution $P_{L,D}(M)$ of the magnetization per spin $M$ in
the system and the hysteresis behavior that one finds in Monte
Carlo simulations applying the single spin flip algorithm
\cite{89} (that realizes the kinetic Ising model with
non-conserved magnetization \cite{90}). If one applies periodic
boundary conditions in both $x,y$ directions, the Hamiltonian of
the model simply is (we take the lattice spacing as our unit of
length in this section)

\begin{eqnarray} \label{eq1}
&&\mathcal{H}=-J \sum\limits_{i=1}^L \, \sum\limits_{j=1}^D \,
S(i,j) [S(i+1,j)+S(i,j+1)] \nonumber\\
&&- H \, \sum\limits_{i=1}^L \, \sum\limits_{j=1}^D S(i,j) \, ,
\end{eqnarray}

where we label the lattice sites by two indices $(i,j)$ in $x,y$
directions, $S(i,j)=\pm 1$, $J$ is the exchange energy, and $H$
the (normalized) magnetic field. Here, we are interested in the
limit $L \rightarrow \infty$ for finite $D$. Note that lengths 
like $L$, $D$, $\xi$ etc.~ are dimensionless in the further analysis. First, we summarize
some exactly known results which are useful for our analysis:

(i) The system does not develop a spontaneous magnetization.
Rather the spin correlation function for large distances $x$ shows
an exponential decay \cite{63,71,77}, for zero magnetic field,

\begin{eqnarray} \label{eq2}
&&g(x) = \langle S(i,j)S(i+x,j)\rangle_T \propto\exp (-x/\xi_D),\  x \rightarrow \infty
\end{eqnarray}

with the correlation length of the strip $\xi_D$ being given by
\cite{63}

\begin{equation} \label{eq3}
\xi^{-1}_D=-\frac{1}{2} \gamma_0 -\frac{1}{2} \,
\sum\limits_{r=1}^{2D-1} (-1)^r \gamma_r \,
\end{equation}

where $(\beta \equiv(k_BT)^{-1})$,

\begin{equation} \label{eq4}
\gamma_0=2 \beta J + \ln \tanh (\beta J)
\end{equation}

and

\begin{equation} \label{eq5}
\cosh \gamma_r= \cosh (2 \beta J) \coth (2 \beta J) - \cos (r
\pi/D) \quad.
\end{equation}

Note that in the limit $D \rightarrow \infty$ we simply get
$\xi^{-1}_\infty = - \gamma_0$. At low temperatures ($\beta J $
large) Eq.~(\ref{eq3}) can be simply approximated by the result
(neglecting logarithmic corrections of order $\ln D$)

\begin{equation} \label{eq6}
\ln \xi_D \approx \beta D \sigma \quad ,
\end{equation}

where $\sigma$ is the interfacial tension of the bulk
two-dimensional Ising model \cite{91}

\begin{equation} \label{eq7}
\sigma=2 J - \beta^{-1} \ln [(1+\exp (-2 \beta J))/(1-\exp (-2
\beta J))] \quad .
\end{equation}

\begin{figure}
\includegraphics[scale=0.35]{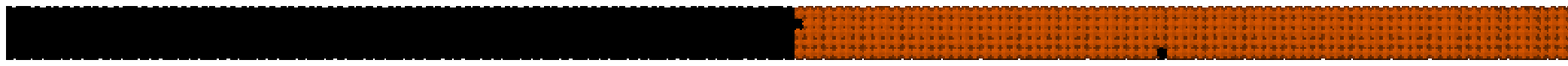}
\includegraphics[scale=0.35]{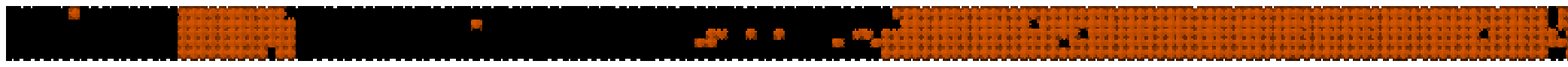}
\includegraphics[scale=0.35]{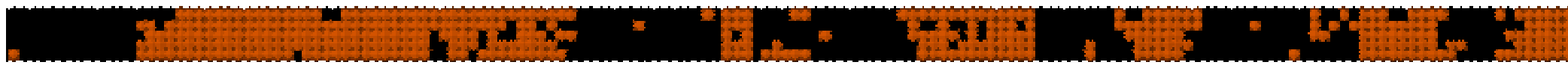}
\includegraphics[scale=0.35]{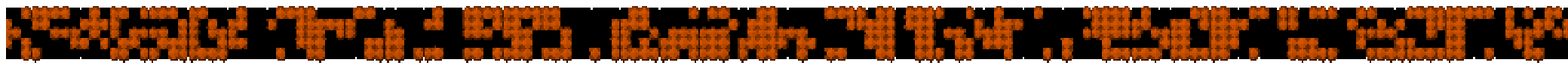}
\caption{Sectors of size 180$\times$5 of an Ising system
with $L=10^5$, $D=5$ at temperatures $T$ (in units of $J/k_B)$
$1.1,1.8,2.2$ and $4.4$ (from top to bottom). Up spins are shown in
black and down spins are shown in gray. For $T=1.1$ the sector was
deliberately chosen such that it contains a domain wall in the
center of the sector. The magnetization is zero for all snapshots.}\label{fig1}
\end{figure}
Eq.~(\ref{eq6}) may simply be understood in terms of a description
of the Ising strip at low temperatures as a (dilute) gas of domain
walls oriented in the $y$-direction and separating large domains
of opposite magnetization \cite{71}. Such a description is
plausible when one looks at snapshot pictures of the Ising strip
(Fig.~\ref{fig1}). Eq.~(\ref{eq6}) simply follows when one asks
at which length $L$ of a quasi-one-dimensional system the free
energy difference $\Delta F$ of a system with a domain wall (of
free energy cost $F_{\rm int}$) and a system in a mono-domain
configuration vanishes, taking the entropy gain $(\ln L)$ of
putting the interface somewhere, into account \cite{61,71}

\begin{equation} \label{eq8}
\Delta F = F_{\rm int} - \beta ^{-1} \ln L \quad , \quad F_{\rm
int} = D \sigma \quad.
\end{equation}

\begin{figure}
\includegraphics[scale=0.6]{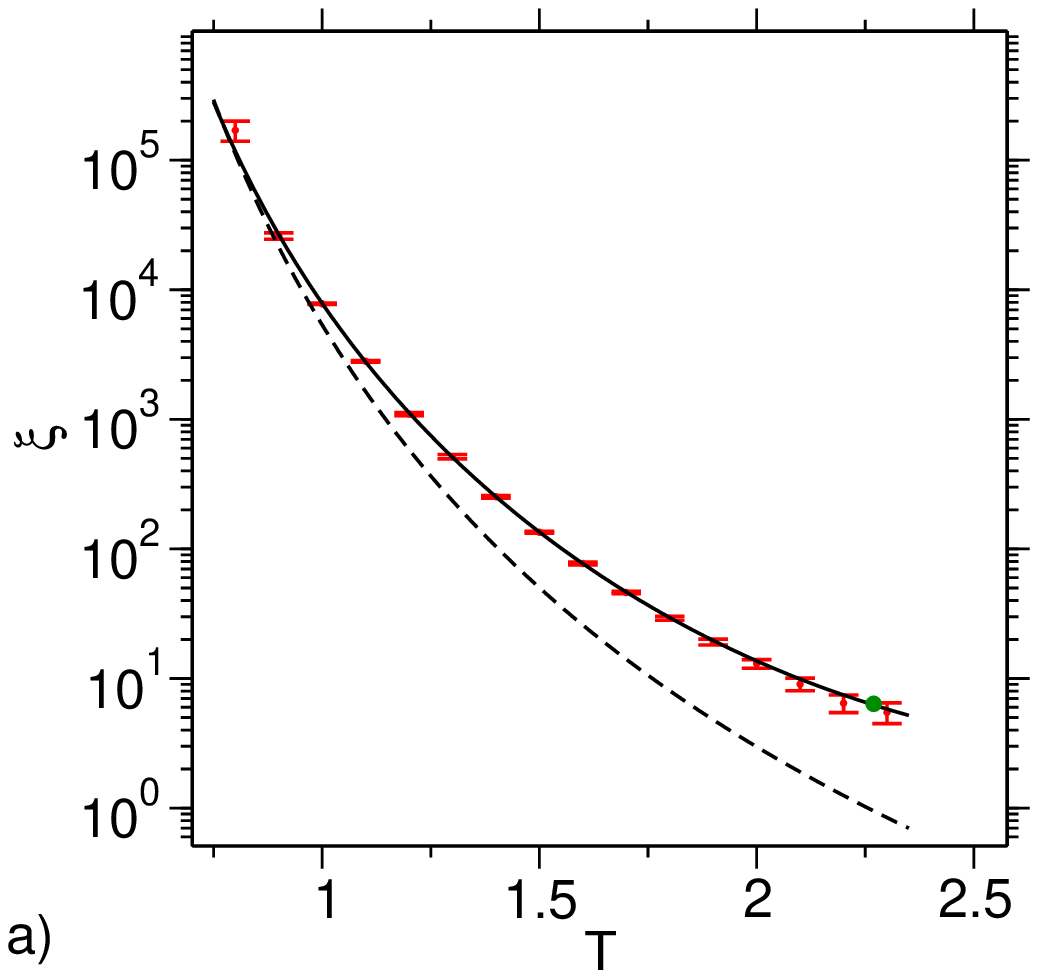}
\includegraphics[scale=0.6]{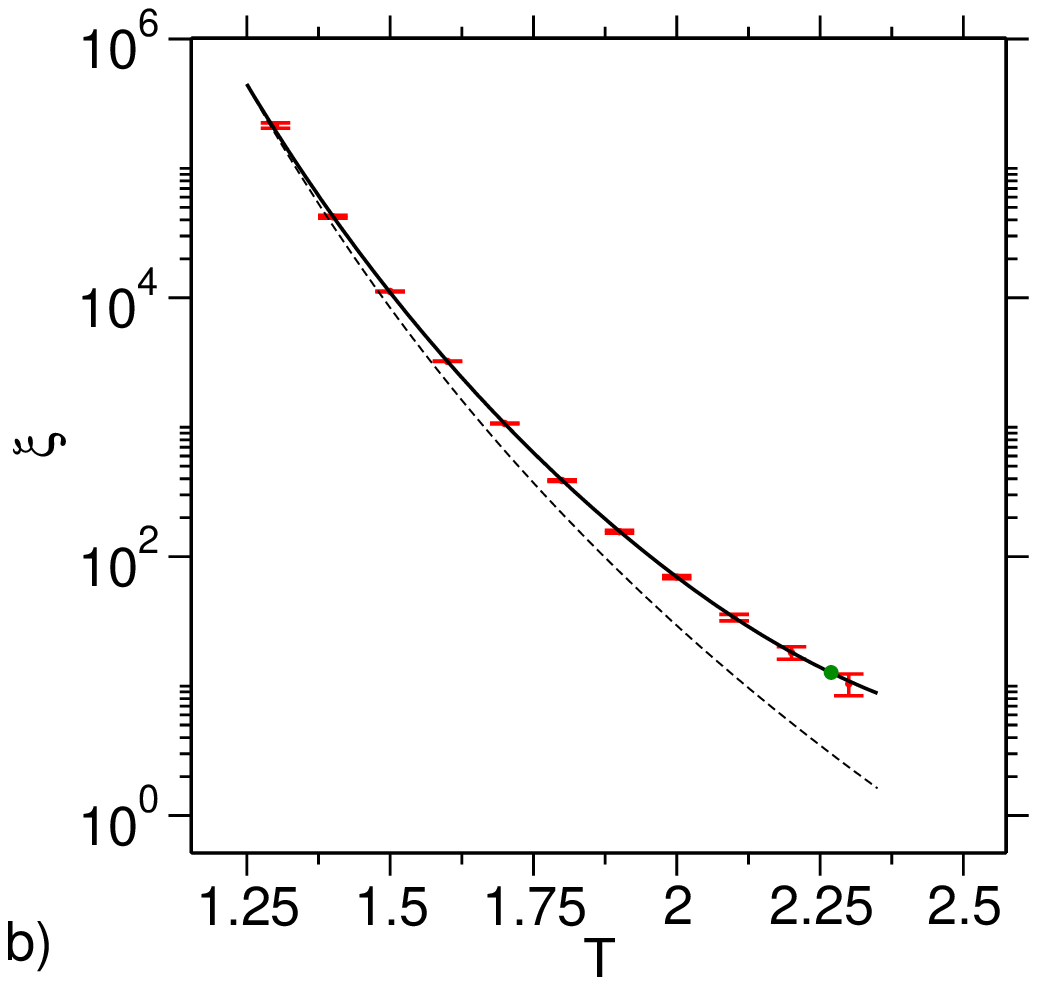}
\includegraphics[scale=0.6]{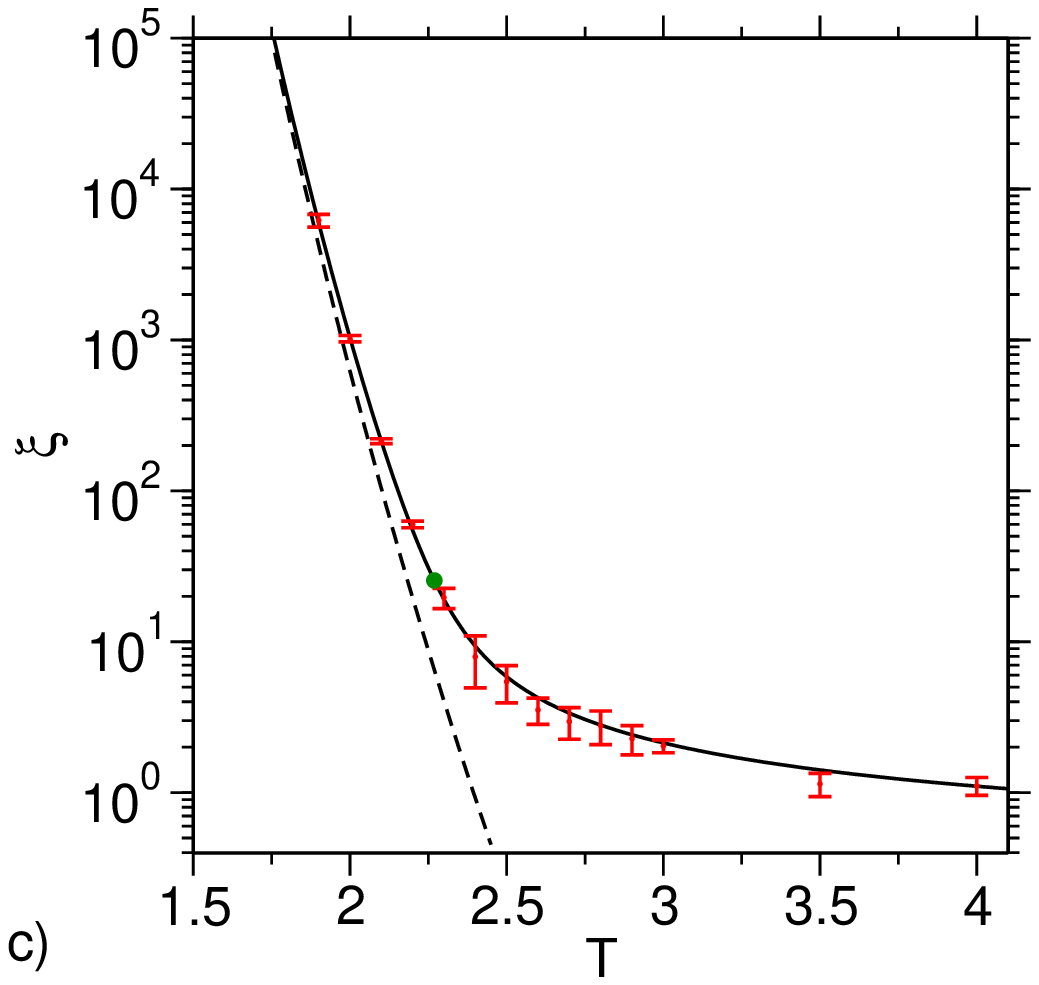}
\caption{Correlation length $\xi_D$ plotted vs. $T$ for $D=5$
(a), $10$ (b) and $20$ (c). Full curve shows Eq.~(\ref{eq3}) while
broken curve shows the approximation Eq.~(\ref{eq6}). The error
bars show Monte Carlo results that were extracted from systems
with $L=10^5$ (cf. text). The dot highlights the correlation
length of the strip at bulk criticality \cite{81},
$\xi_D(T_c)=4D/\pi$. Note the logarithmic scale of the
ordinate.}\label{fig2}
\end{figure}
However, Fig.~\ref{fig2} shows that in the temperature regime that
is of interest for the present paper, $1.0 \leq k_B T /J
\leq k_BT_c/J=2/\ln(\sqrt{2}+1) \approx2.269$, Eq.~(\ref{eq6})
holds only qualitatively, but not quantitatively. The exact
results \{Eqs.~(\ref{eq3})-(\ref{eq5})\} are also very useful to
check that our Monte Carlo algorithm indeed provides a
sufficiently accurate sampling: using the Wolff \cite{92} single
cluster algorithm, systems of size $L=10^5$ were simulated for
$D=5,10$ and $20$. The (second moment) correlation length $\xi$ in
$x$-direction was then obtained from a sampling of the
wave-vector-dependent susceptibility $\chi(\vec{k})$,

\begin{eqnarray} \label{eq9}
&& \chi(\vec{k})=\beta LD \langle |M(\vec{k})|^2\rangle , \nonumber\\&&
M (\vec{k})=(LD)^{-1} \sum\limits_{j,\ell}S (j,\ell) \exp (i \vec{k}
\cdot \vec{r}_{j,\ell}) \quad ,
\end{eqnarray}

orienting $\vec{k}$ in $x$-direction and using the smallest
possible value $|\vec{k}_{\rm min}|=k_{\rm min} = 2 \pi/L$, to
obtain

\begin{equation} \label{eq10}
\xi= \frac{1}{2 \sin (k_{\rm min}/2)}
\Big[\chi(0)/\chi(\vec{k}_{\rm \min})-1\Big]^{1/2} \quad .
\end{equation}

Eqs.~(\ref{eq9}),~(\ref{eq10}) are known as an efficient method to
minimize finite size effects on the estimation of the correlation
length $\xi$ when $L$ and $\xi$ are of comparable size \cite{93},
which is true for the case of the lowest temperatures studied, and
then the direct estimation of $\xi_D$ from the spin correlation
function \{Eq.~(\ref{eq2})\} becomes cumbersome. Fig.~\ref{fig2}
shows that in this way it has been possible to ``measure'' the
growth of the correlation length over 5 decades accurately.

As is well known, the lack of better quantitative agreement
between the exact result \{Eqs.~(\ref{eq3})-(\ref{eq5})\} and the
approximation based on Eqs.~(\ref{eq6})-(\ref{eq8}) can be
attributed to the ``capillary wave'' \cite{12,13,14,15,16,39}
excitations of the interfaces, which lead to an effective
repulsive interaction between neighboring domain walls \cite{94}
leading to a correction to Eq.~(\ref{eq6}), which for large
$D$ gets replaced by \cite{62}

\begin{equation} \label{eq11}
\xi_D \propto D^{1/2} \exp (\beta \sigma D) \quad .
\end{equation}

However, Eq.~(\ref{eq11}) still fails in the vicinity of $T_c$
where one rather finds \cite{81}

\begin{equation} \label{eq12}
\xi_D(T_c)=4D/\pi \quad .
\end{equation}

(ii) When for $T<T_c$ the magnetic field is varied from positive
to negative values the jump from positive $(+M_0)$ to negative
$(-M_0)$ spontaneous magnetization, that would occur in the
two-dimensional bulk, is slightly rounded. One finds \cite{62,77}
$(M=(LD)^{-1} \sum\limits_{j,\ell} S(j, \ell))$

\begin{equation} \label{eq13}
\langle M \rangle =H \Big\{\chi_\infty + D \frac{M^2_0}{k_B T}/\Big[(2 \xi _D)^{-2} + (H\frac{M_0}{k_BT}D)^2 \Big]^{1/2}
\Big\} \, .
\end{equation}

The first term $(\chi_\infty)$ in the curly brackets is just the
susceptibility at phase coexistence in the bulk ($D \rightarrow
\infty$ first, then $H \rightarrow 0)$ for $T<T_c$. The second
term describes the rounding of the transition: it extends over a
region of fields where both terms in the square bracket of the
denominator in Eq.~(\ref{eq13}) are of the same order \cite{77}

\begin{equation} \label{eq14}
H=\pm k_B T / (2M_0 D \xi_D) \propto \pm D ^{-3/2} \exp (-\beta
\sigma D) \quad.
\end{equation}

The maximum value of the susceptibility $\chi=\partial \langle M
\rangle / \partial H$ in the strip can then be readily obtained as

\begin{equation} \label{eq15}
\chi_{\rm max} = \chi_\infty + 2 \frac{M_0^2}{k_BT} D \xi_D\propto
D^{3/2} \exp (\beta \sigma D) \quad .
\end{equation}

As it should be, we find that the region of the rounding
\{Eq.~(\ref{eq14})\} times $\chi_{\rm max}$ covers just the range
$\pm M_0$.

The simple result for $\chi_{\rm max}$ is easily interpreted in
terms of the fluctuation relation (for $L \rightarrow \infty$;
note that in the $\sum_{i, \ell}$ all relative distances occur
twice)

\begin{eqnarray} \label{eq16}
&&k_BT \chi_{\rm max}=LD \langle M^2 \rangle_{H=0}=(LD)^{-1}
\sum\limits_{i,j,\ell,n} \, \langle S(i,j)S(l,n) \rangle
\approx \nonumber\\
&& 2 M^2_0 D \sum\limits_{\ell=0}^\infty \langle S(i,j) S(i+\ell,
j) \rangle \approx \nonumber \\
&& 2 M^2_0 D \int\limits_0^\infty dx \exp [-x/\xi_D]=2M^2_0 D
\xi_D \quad .
\end{eqnarray}

Rather than taking correlations in the $y$-direction exactly into
account, $\sum\limits_{j=1}^D S(i,j)$ is approximated by
$M_0DS(i,j)$ in the first approximation step in Eq.~(\ref{eq16}).
Of course, Eq.~(\ref{eq16}) is not to be used near $T_c$ where
$M_0 \rightarrow 0$.

Being interested in the effects due to the finite length $L$ of
the strip, we can replace $g(x)=\exp(-x/\xi_D)$ by $g(x) = \exp
(-x/\xi_D)+\exp [-(L-x)/\xi_D]$, to account for periodic
boundary conditions. Thus instead of Eq.~(\ref{eq15}) we then
obtain

\begin{equation} \label{eq17}
\chi_{\rm max} =\chi_\infty + 2D \frac{M_0^2}{k_BT} \xi_D
\Big[1-\exp (-L/\xi_D)\Big] .
\end{equation}

Eq.~(\ref{eq17}) is in fact a simple description of the crossover
to the maximum susceptibility in the case of very short strips $(L
\ll \xi_D)$ where the system does not have any interfaces in
$y$-direction at $H=0$ in its typical configuration, and the jump
between $\pm M_0$ is controlled by the total volume $LD$ of the
strip, rounding occurring over

\begin{equation} \label{eq18}
H=\pm k_B T /(2M_0 LD)
\end{equation}

and the susceptibility maximum being

\begin{equation} \label{eq19}
\chi_{\rm max}=\chi_{\rm \infty} + 2 M_0^2 DL/k_BT \quad .
\end{equation}

While some aspects of these results were tested for the equivalent
problem of a quantum Ising chain in a transverse field \cite{77},
we are not aware of a full test of these predictions for the
standard Ising model. Albano et al. \cite{78} studied the finite
size scaling of Ising strips in a $D \times L$ geometry near the
bulk critical temperature, demonstrating scaling properties at
constant aspect ratio $D/L$. Another study \cite{79} considered
capillary condensation in Ising strips with boundary fields, but
considered the shift of the transition only, ignoring the
rounding.

\begin{figure}
\includegraphics[scale=0.475]{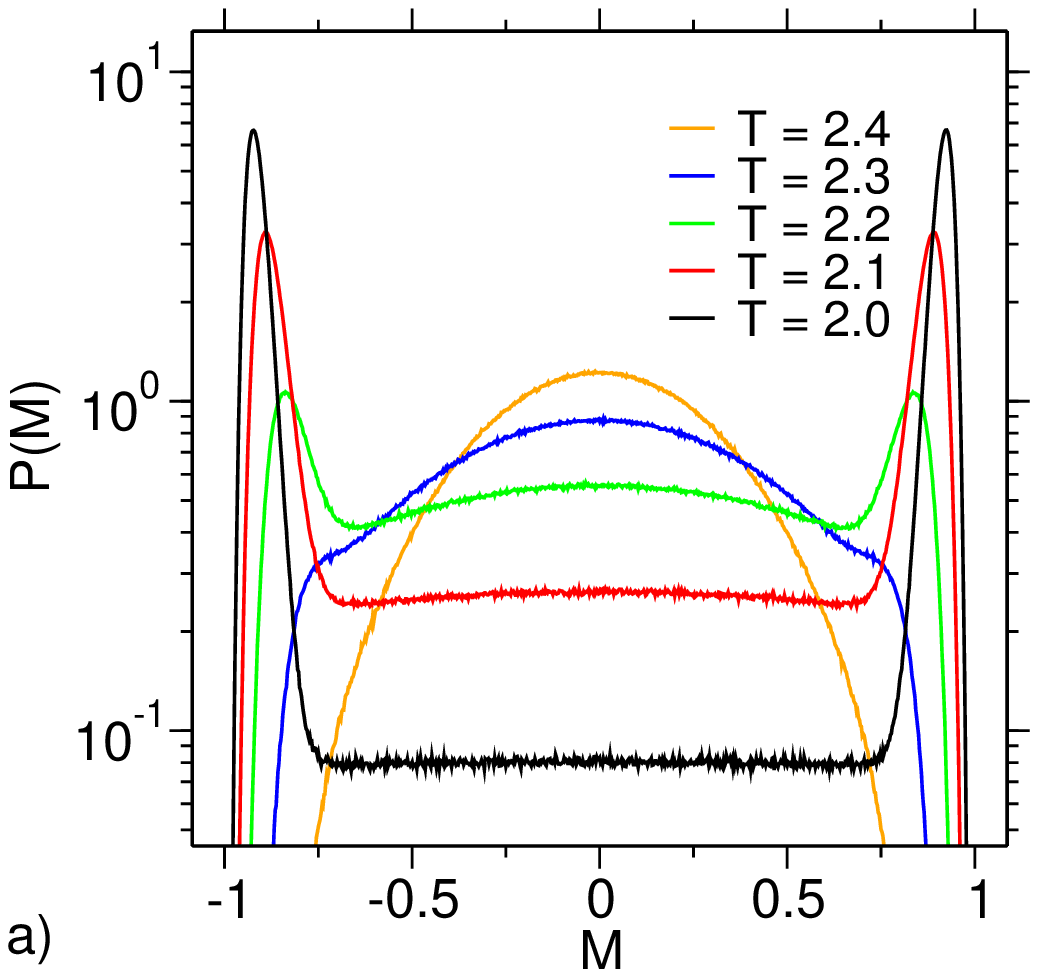}
\includegraphics[scale=0.475]{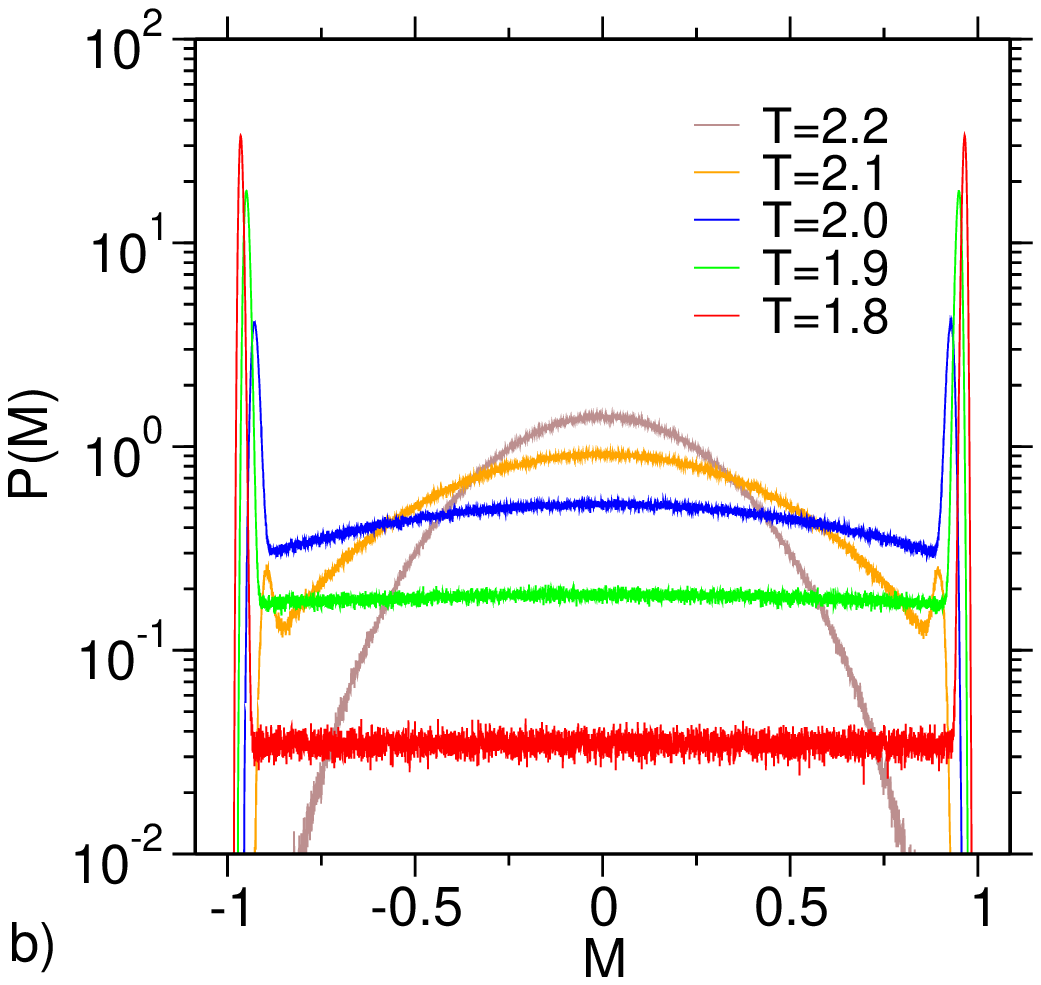}
\includegraphics[scale=0.475]{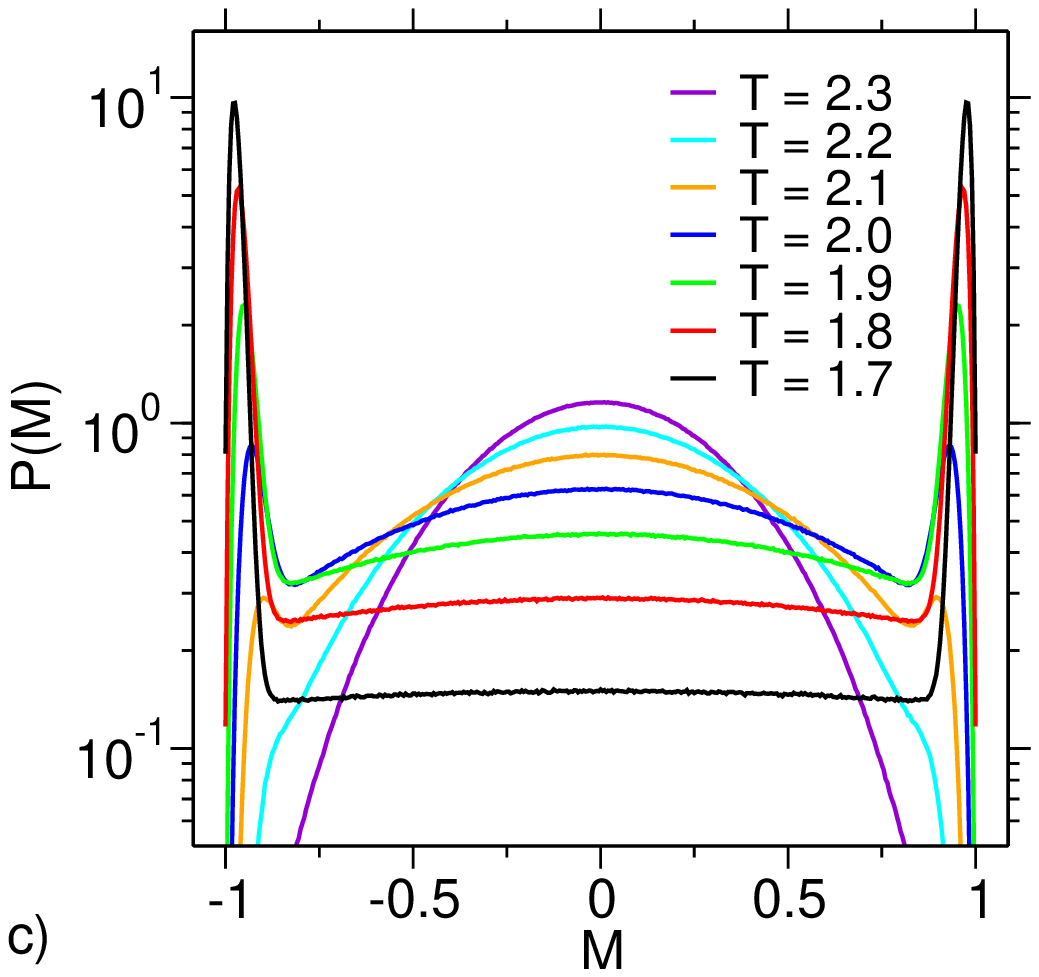}
\includegraphics[scale=0.475]{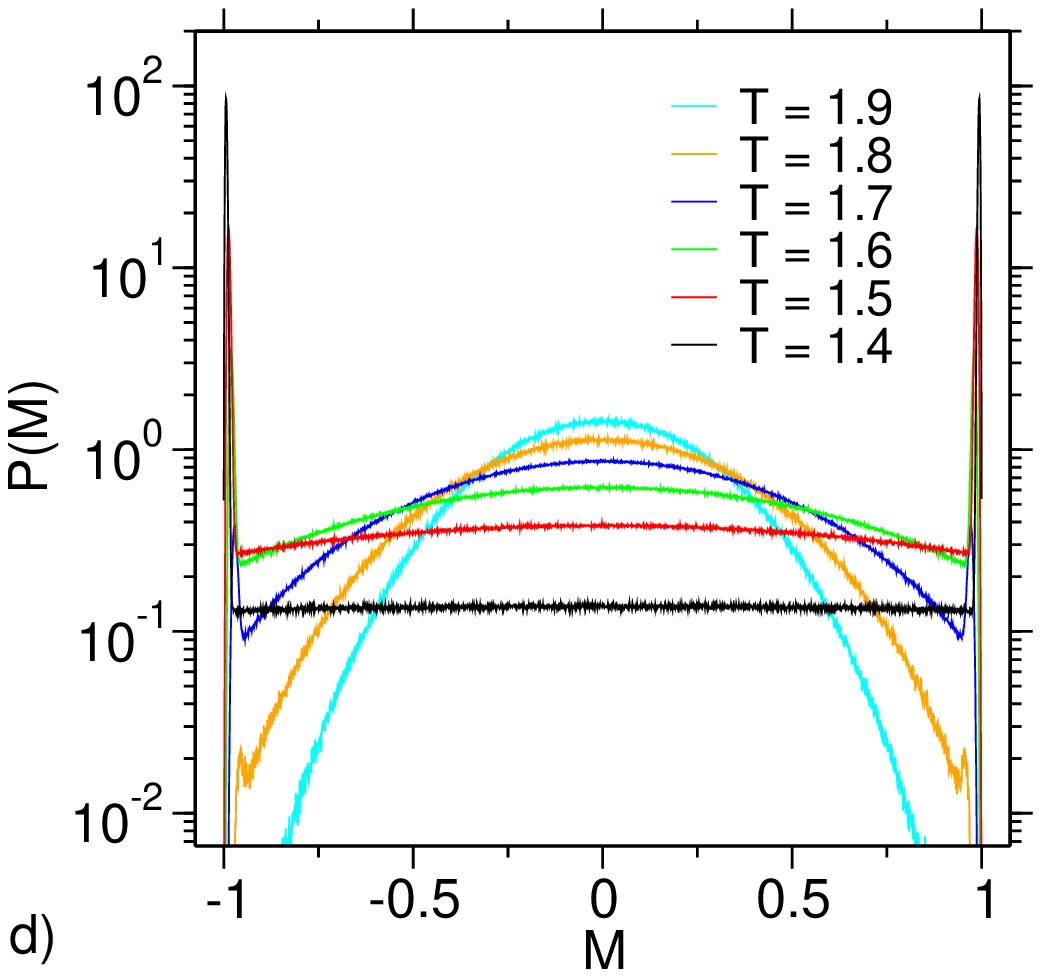}
\caption{Distribution function $P_{L,D}(M)$ plotted vs. $M$ for
a) $D=10$, $L=80$; b) $D=10$, $L=480$; c) $D=5$, $L=80$; d) $D=5$, $L=480$.
At $M=0$ from top to bottom: curves for decreasing temperatures as indicated.}\label{fig3}
\end{figure}
\begin{figure}
\includegraphics[scale=0.6]{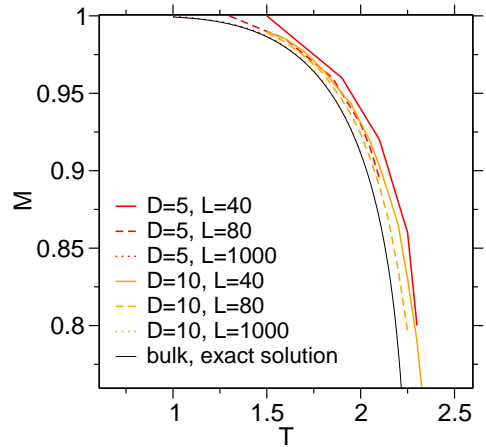}
\caption{Estimates for spontaneous magnetization
extracted from the positions of the peaks of $P_{L,D}(M)$. The continuous curve
shows the exact solution for an infinite system \cite{97}.
Different choices of $L$ and $D$ are indicated.}\label{fig4}
\end{figure}
When one studies phase transitions by Monte Carlo methods
\cite{89}, the standard method of analysis is based on finite size
scaling studies of the order parameter distribution function
$P_{L,D}(M)$ and its moments \cite{81,95,96}. Working at $H=0$, we
 can still use the same cluster algorithm as used for the Monte
Carlo calculation of the correlation length (Fig.~\ref{fig2}), but
now we are also interested in studying the effect of varying both $L$
and $D$ (Fig.~\ref{fig3}). We see that at low temperatures
$P_{L,D}(M)$ has the structure familiar from studies in the
standard square $(L \times L)$ or cube $(L \times L \times L)$
geometry: there rather sharp peaks occur at $\pm M_{\rm max}$
close to $\pm M_0(T)$ (cf. Fig.~\ref{fig4}). For comparison, the
exact solution for the spontaneous magnetization of an infinite
Ising lattice \cite{97} is included. One can see that in this case
finite size effects lead to slightly but systematically larger
values of the magnetization.

At low temperatures, the region of $P_{L,D}(M)$ in between the
peaks has a perfectly horizontal part. As is well known
\cite{96,98}, this flat part is due to the existence of just two,
non-interacting, interfaces crossing the system in $y$-direction.
The free energy cost of creating two interfaces (for $\beta
F_{\rm int} \gg \ln L$ the entropic contribution where the
interfaces at given $M$ are placed, cf. Eq.~(\ref{eq8}), can be
neglected) is simply given by $2 F_{\rm int}=2 D \sigma (T)$, and
actually the estimation of $\ln [P_{L,D} (M_{\rm max})/P_{L,D}(0)]
\approx 2 \beta F_{\rm int}$ is a useful method to
numerically estimate $\sigma (T)$ \cite{96,98}.

However, all the above statements apply only when $L \ll
\xi_D(T)$, and since $\xi_D(T)$ decreases rapidly when $T$
increases (Fig.~\ref{fig2}) the crossover when $L$ and $\xi_D(T)$
are of the same order needs to be considered. In $P_{L,D}(M)$,
this crossover shows up via a three-peak-structure: near $M=0$ a
third peak grows and gains in weight $W$ as $T$ is raised, and
ultimately the peaks near $\pm M_0(T)$ have lost all their weight
and just disappear in the tails of the central peak. In order to
quantify this behavior, we define the weight of the middle peak as

\begin{equation} \label{eq20}
W=\int\limits_{-m}^{+m} P_{L,D}(M) dM / \int\limits^{+1}_{-1}
P_{L,D}(M) d M
\end{equation}

where the minima of $P_{L,D}(M)$ are denoted as $\pm m$. Of
course, at higher temperatures one always reaches a ``spinodal
temperature'' $T_{sp}(L,D)$ where $M_{\rm max}$ and $m$ merge, and
then one no longer has a 3-peak structure, and Eq.~(\ref{eq20})
loses its meaning: however, before this occurs $W$ is practically
indistinguishable from unity. We also emphasize that $T_{sp}(L,D)$
depends on both $L$ and $D$ significantly, and like other
``spinodals'' it does not have any physical significance, for
systems with short-range interactions like considered here
\cite{99}.

\begin{figure}
\includegraphics[scale=0.6]{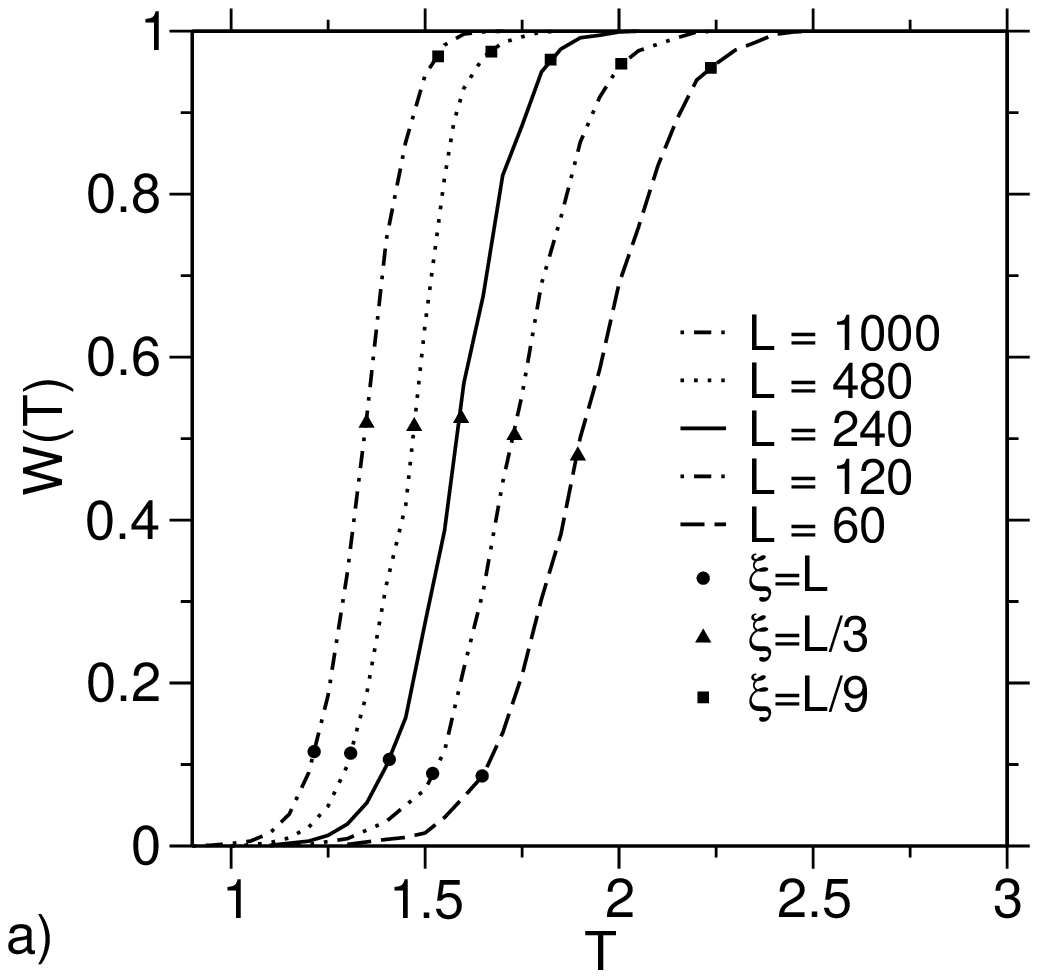}
\includegraphics[scale=0.6]{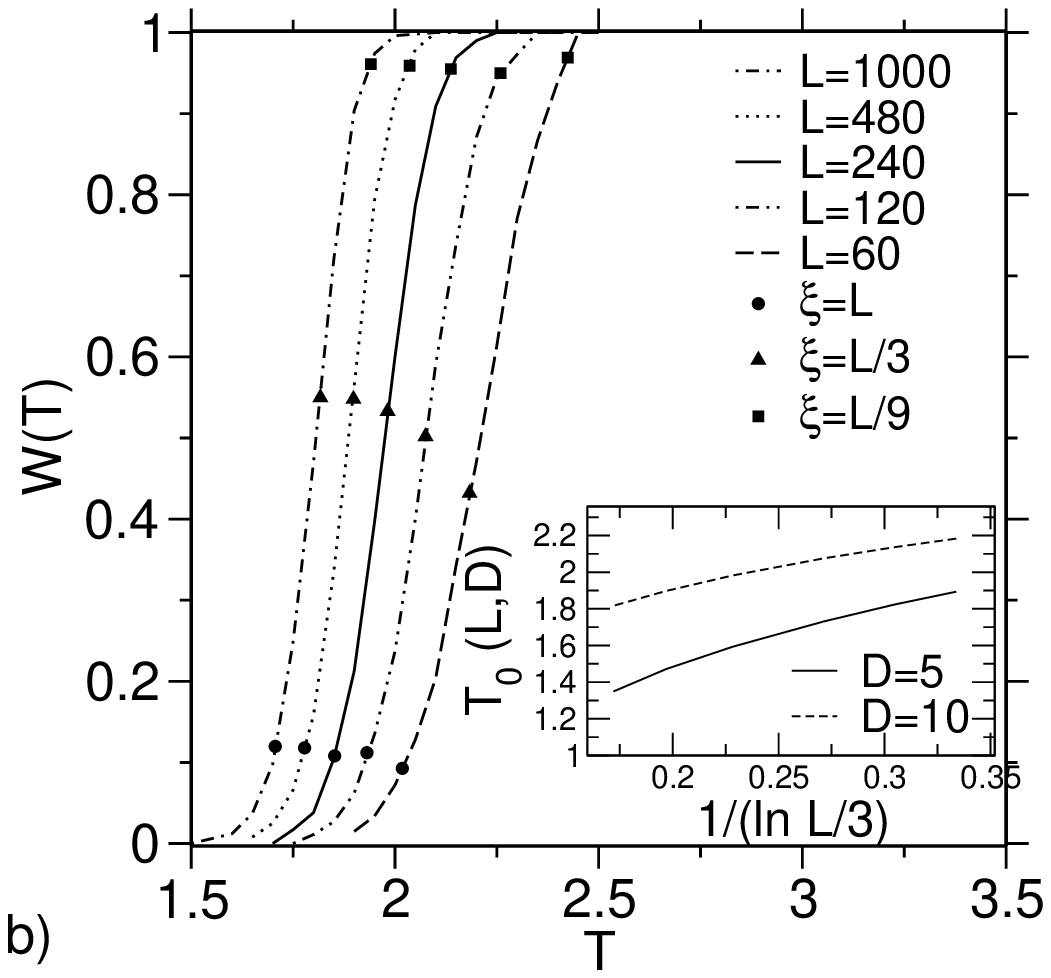}
\caption{Weight $W(T)$ of the central peak of $P_{L,D}(M)$
plotted vs. temperature for $D=5$ (a) and $D=10$ (b). Various
choices of $L$ are included, as indicated. The symbols indicate
the temperatures where $\xi=L$ or $\xi=L/3$ or $\xi=L/9$,
respectively. Insert shows plots of $T_0(L,D)$ vs. $L$,
 cf. text for the definition of $T_0(L,D)$.
}\label{fig5}
\end{figure}
Fig.~\ref{fig5} shows the variation of $W$ with temperature for
two choices of $D$ and a range of values for $L$. We recognize a
gradual increase of $W$ from $W=0$ (two-peak structure with
perfectly flat variation of $P_{L,D}(M)$ near $M=0$) to $W=1$
(single Gaussian peak centered at $M=0$) as $T$ increases.
However, the larger $L$ becomes the more this gradual
transition is depressed to lower temperature, and the sharper it
becomes. It is interesting to correlate this transition with the
fact that $\xi_D(T)$ decreases from values where $\xi_D(T)$
exceeds $L$ to values where $\xi_D(T)$ is much smaller than $L$.
Thus, we have marked three temperatures for each curve where
$\xi_D(T)=L$ ($W$ is close to $0.1$ there) and where $\xi_D(T)=L/3$
($W$ is close to $0.5$ there, i.e.~we are in the center of this
transition region) and where $\xi_D(T)=L/9$ ($W$ is close to
$0.9$ there, i.e. the transition is essentially completed). Thus, we can
define a transition temperature $T_0(L,D)$ where at $H=0$ the
strip experiences a transition from a state where it is typically
ordered $(\pm M_0)$ to a state where it is typically not uniformly
ordered $(\langle M \rangle$ close to $M=0$) although it is
locally ordered (because the system is split into many domains
of typical length $\xi_D(T) \ll L)$. Hence we define $T_0(D,L)$
implicitly via

\begin{equation} \label{eq21}
\xi_D(T_0(L,D))=L/3 \quad ,
\end{equation}

and we define the temperature width $\Delta T$ of this transition 
in terms of

\begin{equation} \label{eq22}
\xi_D (T_0 (L,D)-\Delta T)/\xi_D (T_0(L,D)+\Delta T)=L/(L/9)=9
\quad . \end{equation}

At large enough $L$, where $T_0(L,D)$ is so low that
Eq.~(\ref{eq6}) is accurate, we can use
Eqs.~(\ref{eq6}),~(\ref{eq7}) to rewrite $\xi_D(T)$ as
($X(T)\equiv[\exp (2\beta)-1]/[\exp(-2
\beta)+1)]$, choosing henceforth units where $J/k_B\equiv 1$)

\begin{equation} \label{eq23}
\xi_D(T) =[X(T)]^D \,
\end{equation}

and hence Eqs.~(\ref{eq21}),~(\ref{eq22}) can be rearranged as,
for $L \rightarrow \infty$ ($k_B\equiv 1, J\equiv 1$):

\begin{eqnarray} \label{eq24}
&& X(T_0 (L,D))=\exp \Big[\frac{1}{D} \ln (L/3) \Big],  \nonumber\\
&& T_0 (L,D) \approx 2 D/[\ln (L/3)] \quad .
\end{eqnarray}

Similarly, in this limit the width $\Delta T$ becomes

\begin{equation} \label{eq25}
\Delta T / T_0 (L,D) \approx \ln 3 / \ln (L/9) \quad.
\end{equation}

Thus for $L \rightarrow \infty$ this transition temperature goes
to zero, and the transition becomes gradually sharper and sharper,
but the variations (Eqs.~(\ref{eq24}),~(\ref{eq25})) both are of
order $1/\ln L$ and hence very slow. The inset of Fig.~\ref{fig5}b shows that for the temperatures accessible for our study, Eq.~(\ref{eq24}) is not yet accurate.

\begin{figure}
\includegraphics[scale=0.475]{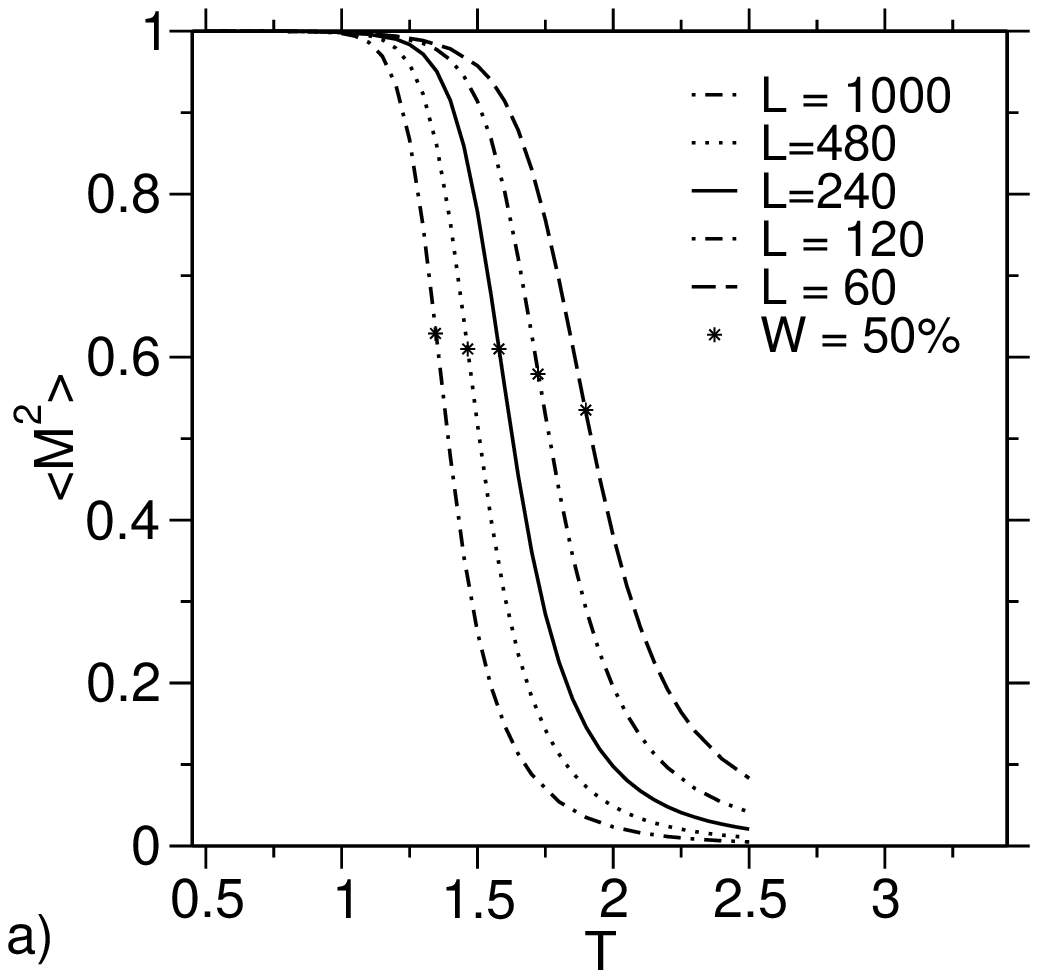}
\includegraphics[scale=0.475]{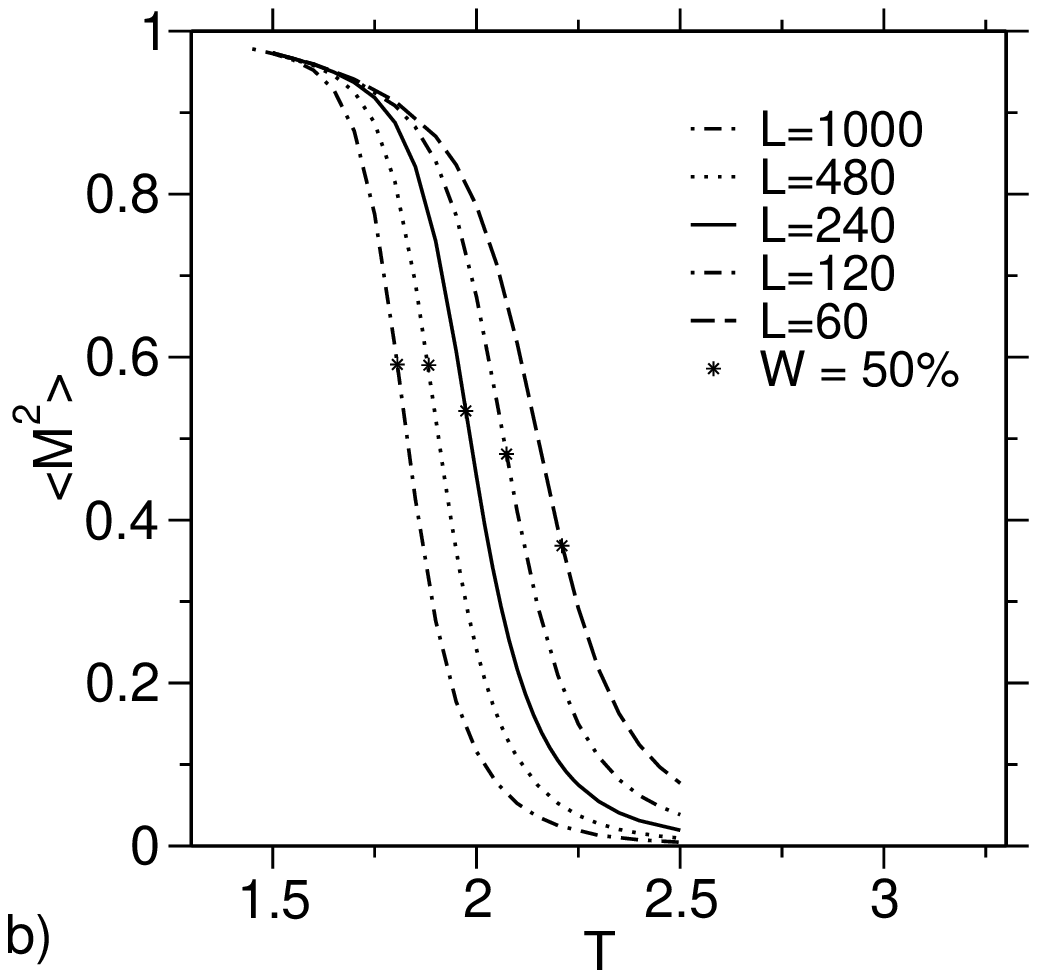}
\includegraphics[scale=0.475]{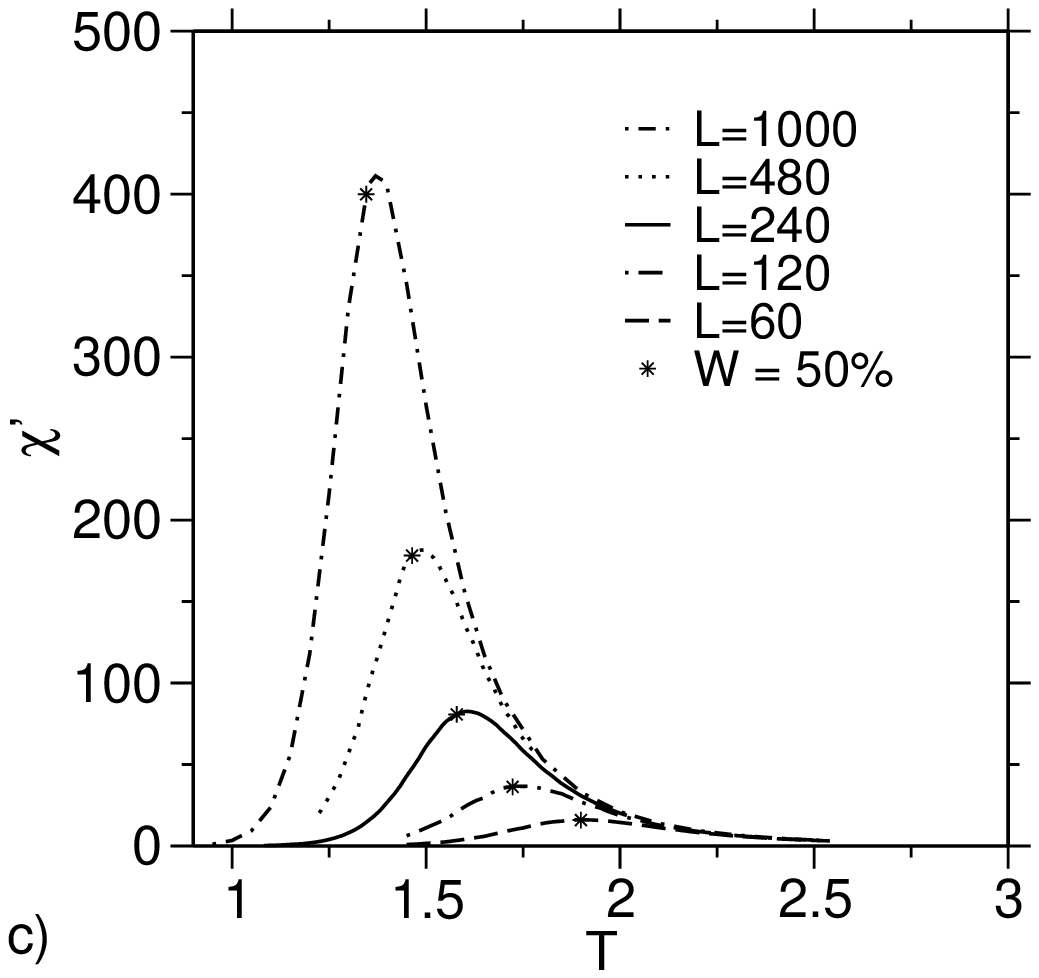}
\includegraphics[scale=0.475]{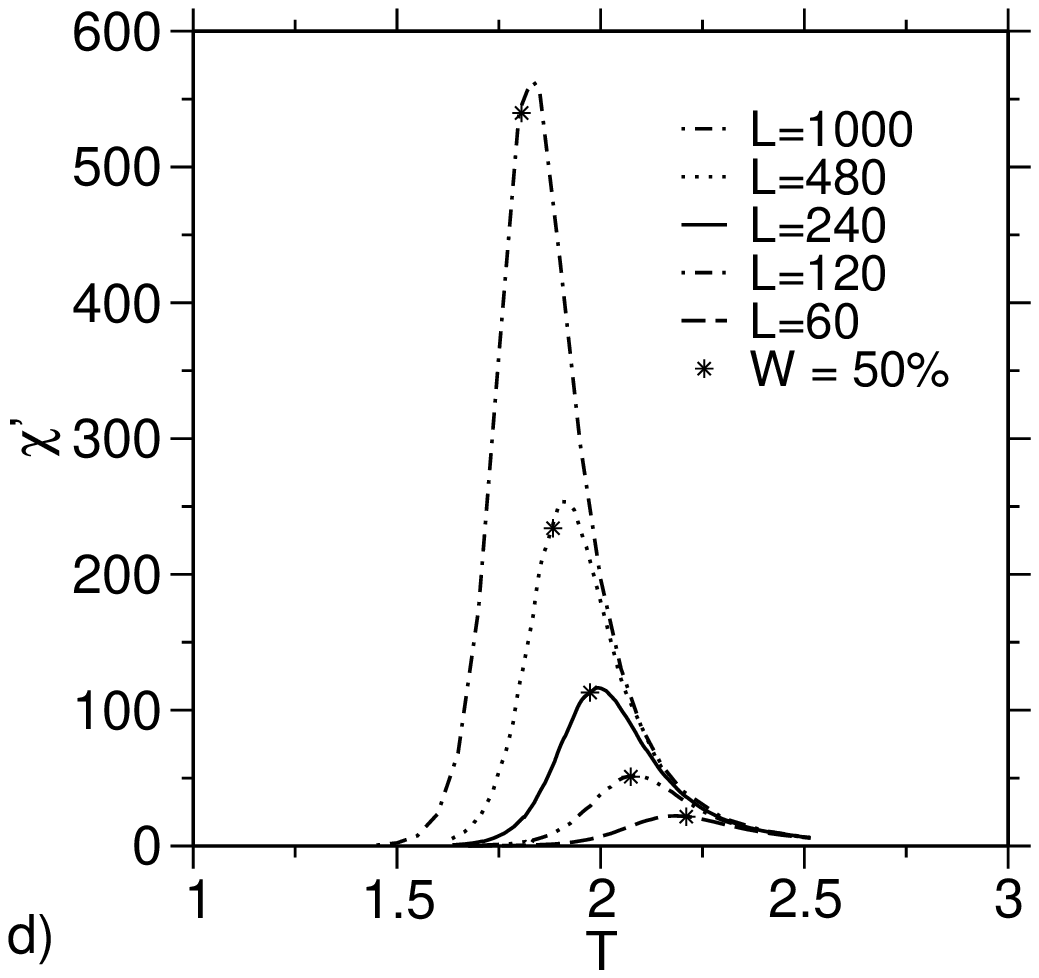}
\caption{Plot of $\langle M^2 \rangle$ (a,b) and $\chi(c,d)$
vs. temperature, for a range of values of $L$, as indicated. Cases
(a,c) refer to $D=5$, cases (b,d) to $D=10$. The asterisk in each
curves marks the temperature at which $W=0.5$.}\label{fig6}
\end{figure}

It is possible to monitor this transition also in a more
conventional way, recording either the temperature variation of
the second moment $\langle M^2 \rangle$ or the ``susceptibility'',
cf. Fig.~\ref{fig6}

\begin{equation} \label{eq26}
\chi'=\beta LD(\langle M^2 \rangle - \langle |M|\rangle^2 )\quad .
\end{equation}

The use of Eq.~(\ref{eq26}) as an estimate for a
``susceptibility'' needs comment: of course, general statistical
mechanics implies that $\chi= (\partial \langle M \rangle
/\partial H)_T=\beta LD(\langle M^2 \rangle - \langle M \rangle ^2
)$, so there does not appear any term involving the absolute value
of the magnetization, and since for $H=0$ we also have $\langle M
\rangle=0$, $\chi$ decreases monotonously with decreasing
temperature, and no maximum occurs. As long as $L \gg\xi_D(T)$,
$\langle |M|\rangle$ is small $(\langle |M|\rangle \rightarrow 0$
for $L \rightarrow \infty$), and then $\chi$ as defined
in Eq.~(\ref{eq26}) differs from the correct susceptibility by a
constant factor (namely $1-2/\pi)$. However, when $P_{L,D}(M)$ for
$L<\xi_D(T)$ just exhibits only two peaks at $\pm M_{\rm max}$, we
have $\langle |M|\rangle \approx M_{\rm max}$ while still $\langle
M \rangle =0$ because of the symmetry of the distribution against
a sign change of $M$. Then $\chi'$ as defined in Eq.~(\ref{eq26})
measures the width of the two peaks of the distribution at $\pm
M_{\rm max}$, while $\chi \approx\beta LDM^2_{\rm max}$. Thus the
peak of $\chi '$ is suitable to give information where the
transition from the multiple domain states at $H=0$ to single
domain states in a finite strip occurs.

As expected, both the peak positions of $\chi'$ and the inflection
points of $\langle M^2 \rangle$ correlate nicely with the
criterion that $W=0.5$. The strong depression of this transition
with increasing $L$ is clearly seen.
\begin{figure}
\includegraphics[scale=0.6]{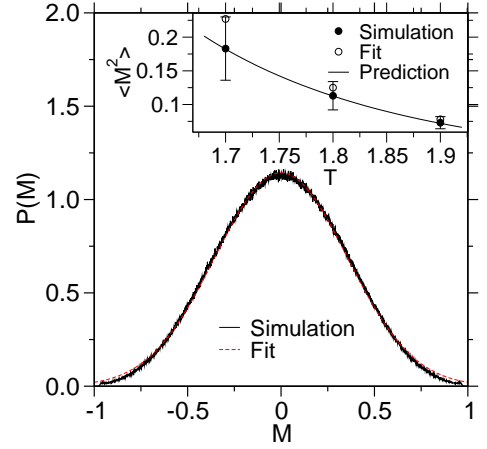}
\caption{Distribution $P_{L,D}(M)$ in the region where
$W \approx 1$ but $T$ is still distinctly smaller than $T_c$, so
a well-identifiable multi-domain configuration is observed. Broken
curves show fit to Eq.~(\ref{eq27}). Insert compares the fitted
value to the prediction, Eq.~(\ref{eq28})} \label{fig7}
\end{figure}

For $T > T_0 (L,D)$ the peaks of $P_{L,D}(M)$ at $\pm M_{\rm max}$
have disappeared, and a broad peak near $M=0$ remains. One can
verify (Fig.~\ref{fig7}) that this peak is simply a Gaussian,

\begin{equation} \label{eq27}
P_{L,D}(M) \propto\exp [-M^2/2 \langle M^2 \rangle ] \quad.
\end{equation}

Noting that \{cf. Eq.~(\ref{eq16})\} $\langle M^2 \rangle = k_BT
\chi _{\rm max}/LD$ we find in this region that

\begin{equation} \label{eq28}
\langle M ^2 \rangle = k_BT \chi_\infty / LD + 2 M^2_0 (\xi_D/L)
\quad .
\end{equation}

The simulations are roughly compatible with this prediction (inset of Fig.~\ref{fig7}).
Finally, we draw attention to the temperature variation of the
free energy barrier between the maximum of $P_{L,D}(M)$ at $M_{\rm
max}$ and the minimum at $m$,

\begin{equation} \label{eq29}
\Delta F= k_BT \ln [P_{L,D} (M_{\rm max})/P_{L,D}(m)] \quad.
\end{equation}

\begin{figure}
\includegraphics[scale=0.6]{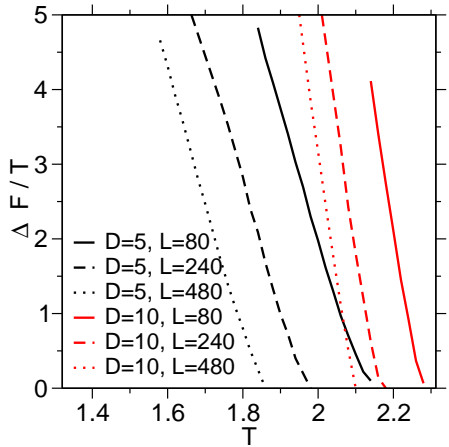}
\caption{Barrier $\Delta F$ against nucleation of interfaces in
Ising strips plotted vs. temperature. Several choices of $L$ and
$D$ are shown as indicated. The three rightmost curves correspond to $D=10$, the three leftmost curves to $D=5$. In each case, from left to right: $L=480$, $240$, $80$.}\label{fig8}
\end{figure}
\begin{figure} 
\includegraphics[scale=0.6]{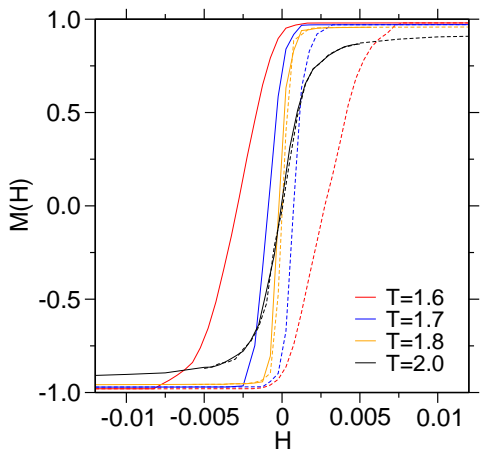}
\caption{Magnetization of Ising strips for $L=480$, $D=10$ plotted vs.
field $H$ at $T=1.6, 1.7, 1.8$ and $2.0$. Runs with
decreasing $H$ are shown as full curves, runs with increasing $H$
as broken curves. While for $T=1.6$ a strong hysteresis can still be observed, a detailed analysis shows that hysteresis
disappears at $T= 1.9$.}\label{fig9}
\end{figure}
Fig.~\ref{fig8} shows a plot of $\Delta F/T$ vs. $T$ for various
choices of $L$ and $D$. The temperatures where these barriers
extrapolate to zero would define the ``spinodal temperatures''
$T_{sp}(D,L)$ already mentioned above, but this is not the point
we want to make now: rather we emphasize that barriers $\Delta F
\approx 10 k_BT$ are reached at temperatures far below $T_c$,
where the local magnetization within a domain (Fig.~\ref{fig4})
is still large. When $\Delta F$ becomes of the order of $10k_BT$
or less, nucleation of domain walls becomes easy, when $H$ is
decreased and one wants to reverse the magnetization in the
system. To test this consideration, we have performed computations
of the magnetization reversal process of the Ising strips, using
the single spin flip Monte Carlo algorithm \cite{89} to realize a
(physically at least qualitatively realistic) dynamical evolution
of the system (in terms of the Kinetic Ising model \cite{90}).
Starting out at $H=0.05$ we decrease the magnetic field in steps
of $\Delta H=0.001$, equilibrating at each state for $\Delta t=2$ million
Monte Carlo steps per spin. Fig.~\ref{fig9} shows some examples of the
hysteresis loops that were recorded in this way. As expected,
hysteresis loops become quickly narrow as the temperature is
increased and when $\Delta F \approx 10 k_BT$ hysteresis
essentially disappears completely. As a consequence, we see that
the ``hysteresis critical temperature'' $T_{ch}$, where hysteresis
loops of our strips disappear, has nothing to do with $T_c$. (It also does not have anything to do with a
finite size analog of $T_c$, where no longer distinct domains of
opposite magnetization in the strip can be distinguished, but one
has more or less isotropic clusters of correlated spins of size
$\xi _D \approx D$\,!). Instead, it correlates rather well with
$T_0(L,D)$, the temperature where no longer uniformly ordered
domains (over the full length $L$ of the strip) are stable.

\subsection{Ising cylinders without surface fields}

Now we consider the analog of Eq.~(\ref{eq1}) on the simple cubic
rather than the square lattice, but remove all lattice sites
with $x$ and $y$ coordinates (when we define the $z$-axis as the
axis of the cylinder) that satisfy

\begin{equation} \label{eq30}
x^2 + y^2 > R^2 \quad .
\end{equation}

As a boundary condition, we first choose the simple free boundary
condition, i.e. interactions to ``missing spins'' do not occur. Of
course, due to the lattice structure (which does not fit to a
cylindrical surface) we have necessarily inequivalent sites at the
surface: i.e., for $R=4$ any cross section of the ``cylinder'' is
not a sphere bounded by a circle, but rather we have 4 spins with
three missing neighbors each (in a positive and negative $x,y$
directions), on next nearest neighbor sites to those sites we have
8 spins with one missing neighbor, and then 8 spins with 2 missing
neighbors follow. The consequence of this non-uniformity of the
boundary condition have not been studied, however, since we do not
consider it to be of real physical interest.

\begin{figure} 
\includegraphics[scale=0.6]{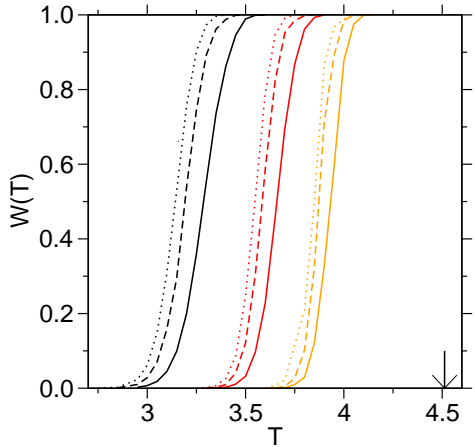}
\caption{Weight of the central peak $W$ \{Eq.~(\ref{eq20})\} of the
order parameter distribution for three choices of the radius $R$ ($R=3$, $4$, $5$ from left to right,
in different colors)
 and three choices of $L$ in each case: $L=600$, $400$, $200$ from left to right with different line styles. The
arrow shows the critical temperature $T_c$ \cite{102} of the bulk
three-dimensional Ising model. Note that a periodic boundary
condition is only applied along the $z$-axis of the ``cylinder''.}
\label{fig10}
\end{figure}
\begin{figure}
\includegraphics[scale=0.475]{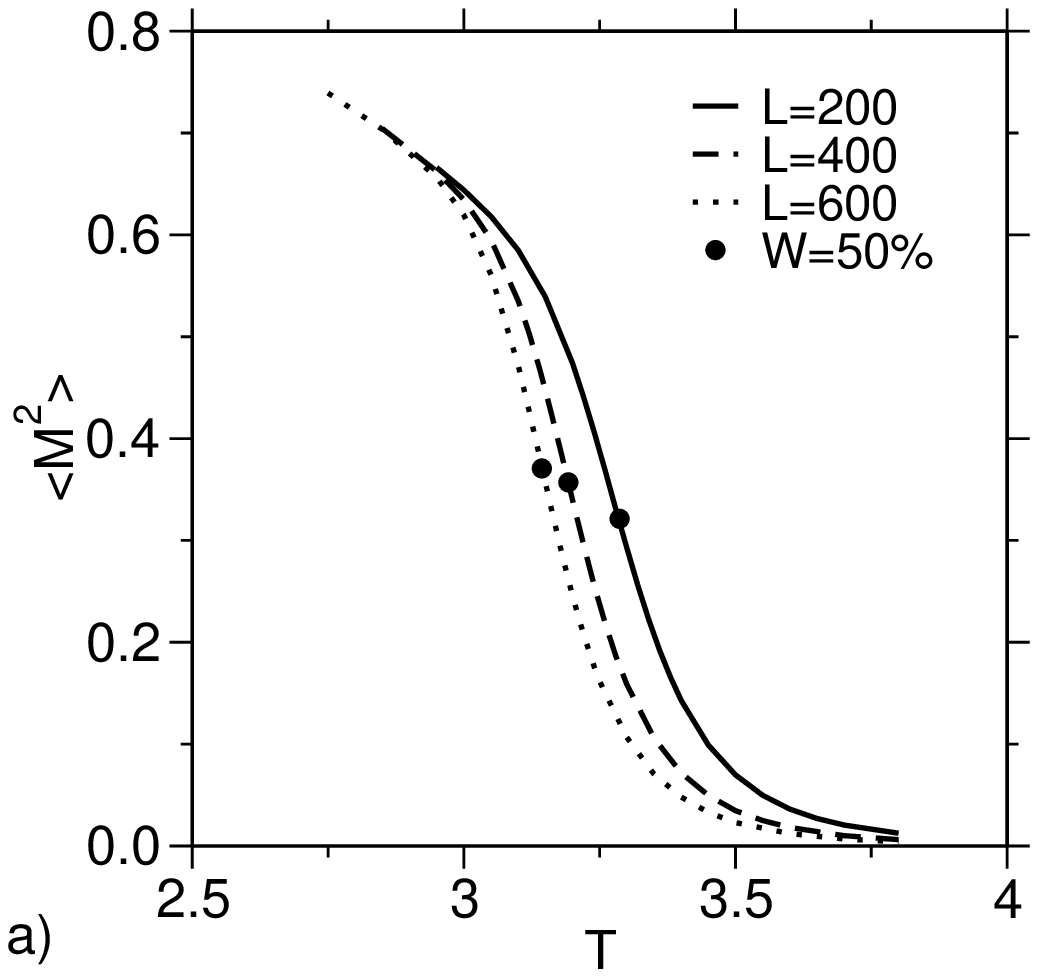}
\includegraphics[scale=0.475]{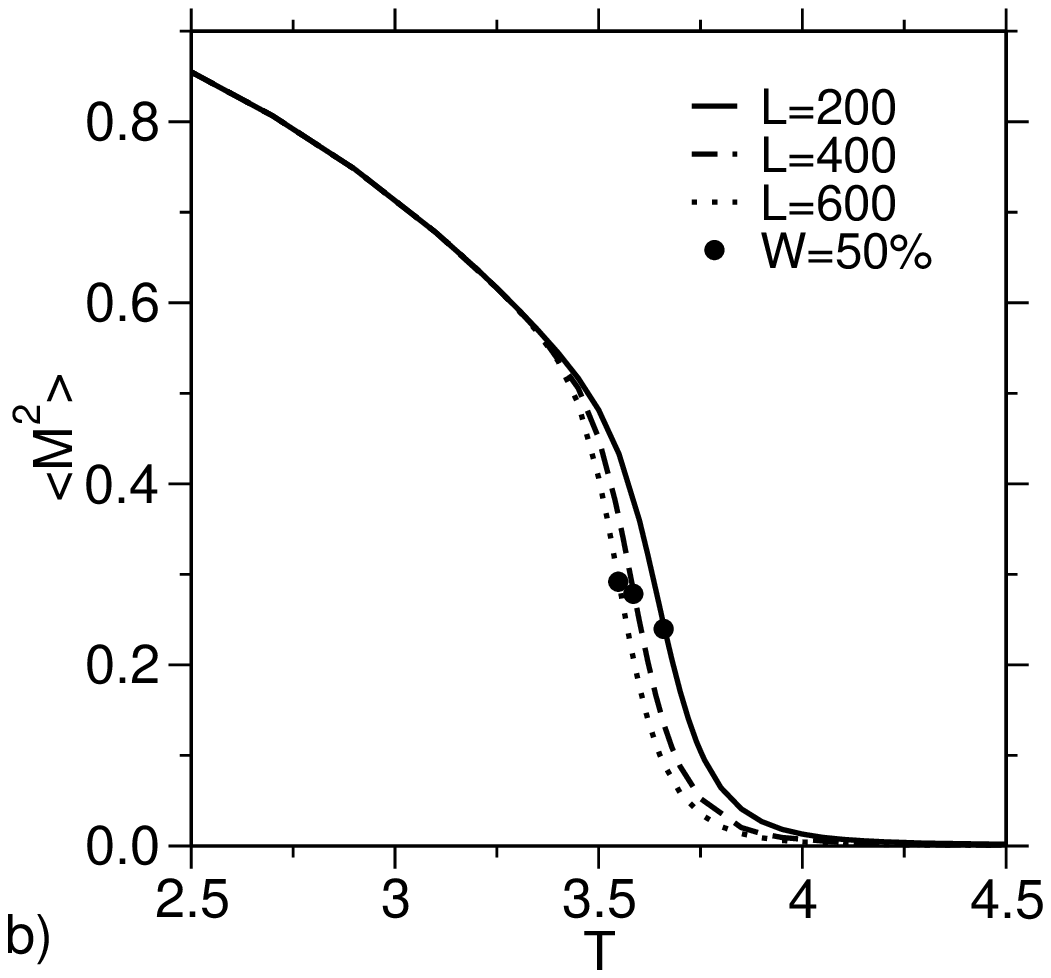}
\includegraphics[scale=0.475]{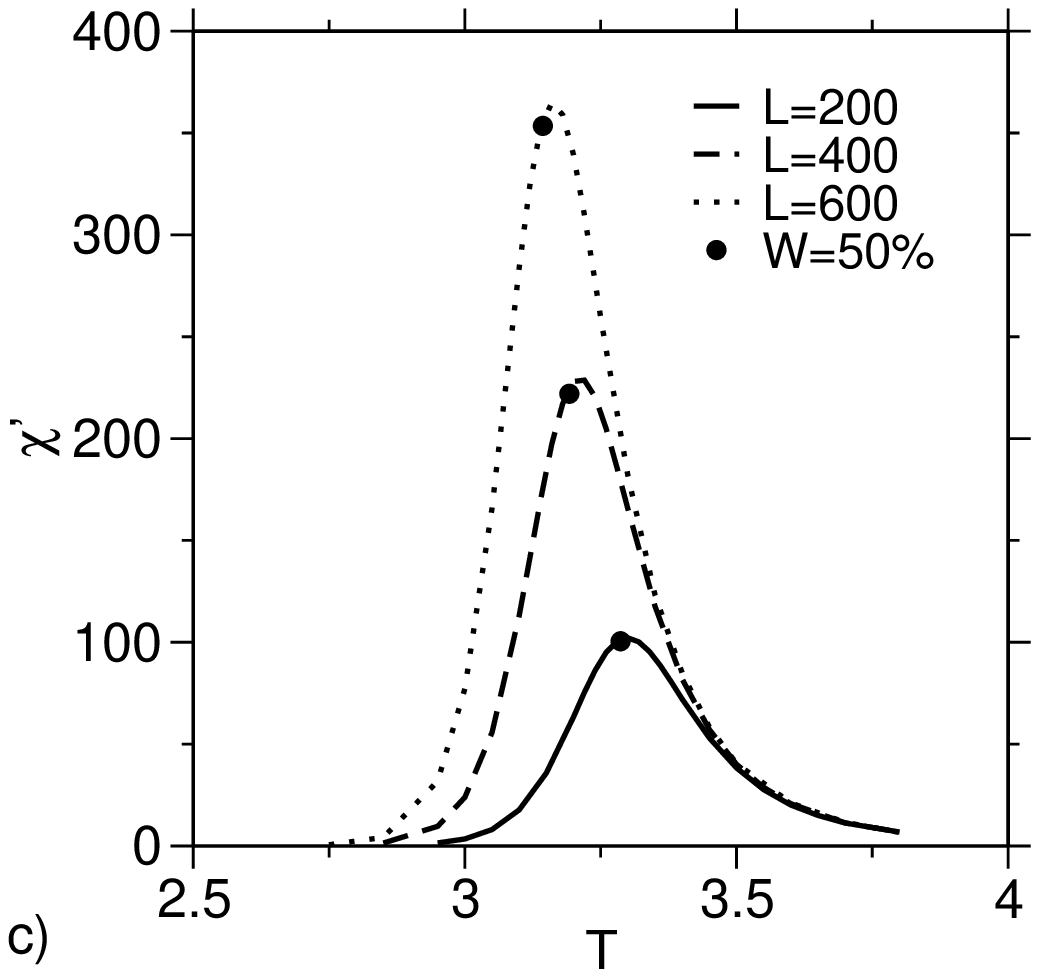}
\includegraphics[scale=0.475]{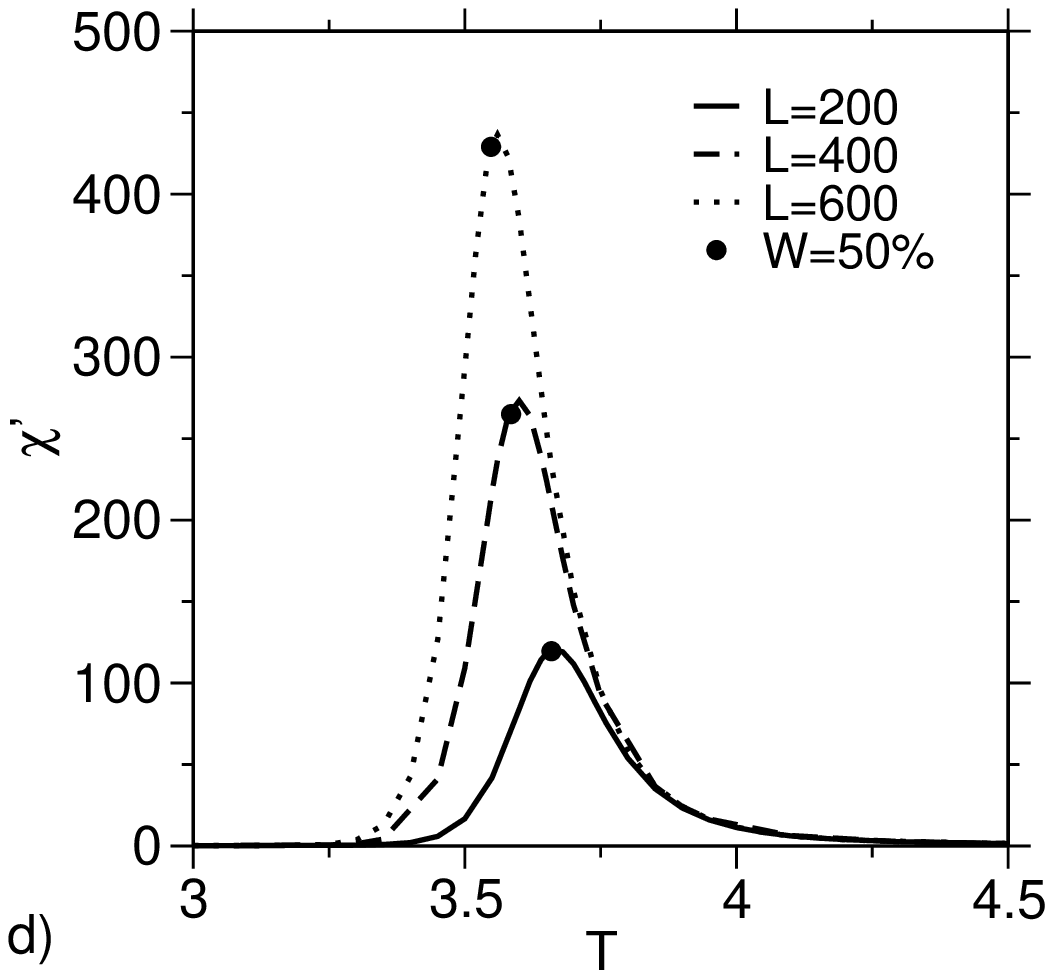}
\caption{Plot of $\langle M^2 \rangle$ (a,b) and $\chi' (c,d)$
 vs. temperature, for a range of values of $L$, as indicated. Cases (a,c) refer to
 $R=3$ and cases (b,d) to $R=4$. Dot on each curve shows the temperature for
 which $W=0.5$.}\label{fig11}
\end{figure}
\begin{figure}
\includegraphics[scale=0.6]{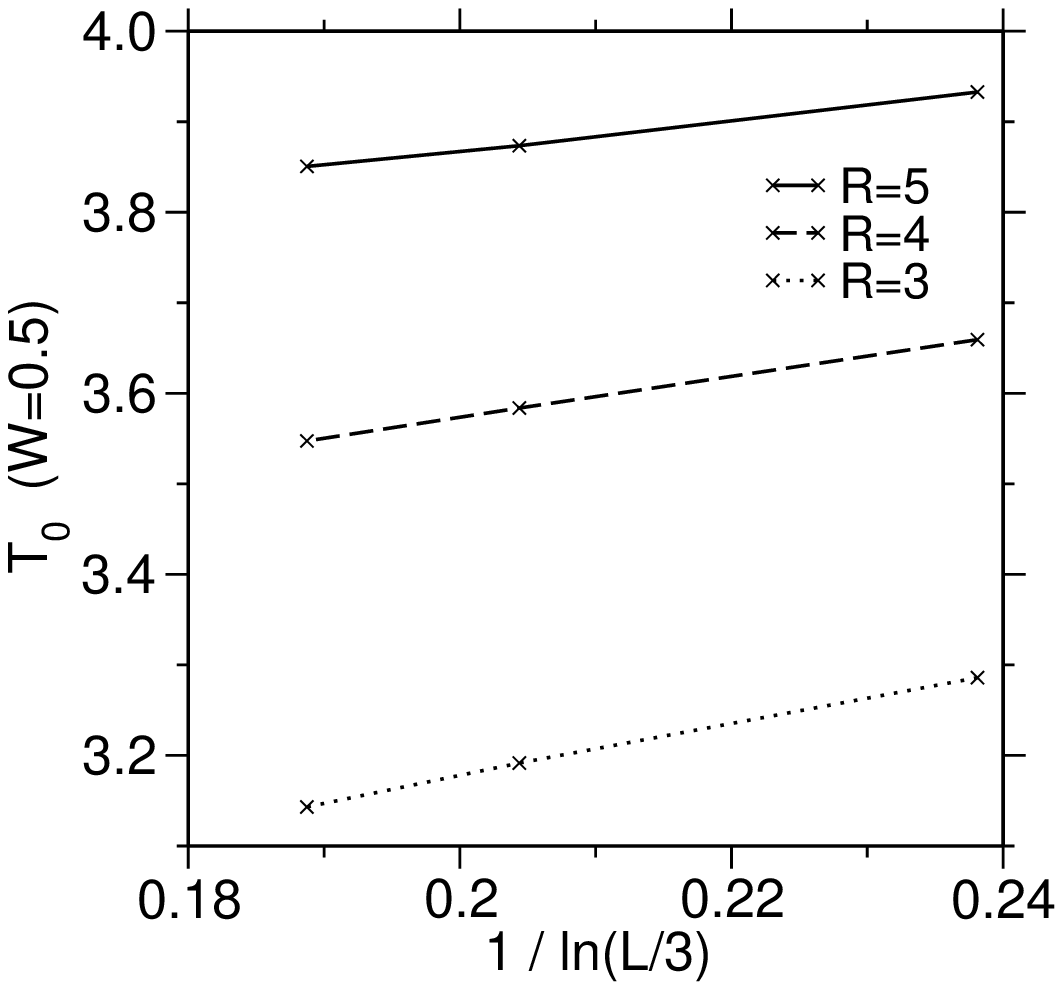}
\caption{$T_0(L,D)$ for three choices of $R$ (as indicated)
plotted vs. $[\ln (L/3)]^{-1}$}\label{fig12}
\end{figure}
If one does not apply any ``surface magnetic fields'' $H_1$
\cite{100,101} at these boundaries the single cluster algorithm
\cite{92} can be straightforwardly implemented for this problem as
well, and the probability distribution $P_{L,D}(M)$ (where $D=2 R$
is the diameter of the cylinder) and its moments can be recorded,
as described in Sec.~\ref{Theory}. Fig.~\ref{fig10}-\ref{fig12} show that
the findings indeed are qualitatively similar. Of course, $\chi'$
in $d=3$ dimensions is defined as (for $H=0$)

\begin{equation} \label{eq31}
\chi' = \beta N (\langle M^2 \rangle -\langle |M|\rangle^2)
\end{equation}

where $N$ is the total number of spins belonging to the
``cylinder''. Unlike the two-dimensional strips (with
periodic boundary conditions also in the $y$-direction across the
strip) now the ordering tendency is strongly suppressed already
for the smallest value of $L$ and due to the missing spins at the
boundary the order in the ``cylinder'' is destabilized as expected.
The same effect is seen when one studies $L \times L$ squares or
$L \times L \times L$ cubes or $L \times L \times D$ films with
free boundaries, as is well known \cite{103,104,105,106}.

We also expect in this case a simple exponential decay of the spin
correlation function in the axial direction of the cylinder,
analogous to Eq.~(\ref{eq2}), but the corresponding correlation
length $\xi_D$ is not known independently. In analogy to
Eq.~(\ref{eq6}), we expect that $\xi_D$ at low temperatures (and
$L \rightarrow \infty$) simply varies exponentially with the
cross-sectional area $A$ of the cylinder,

\begin{equation} \label{eq32}
\xi_D \propto\exp (\beta A \sigma), \quad A=D^2 \pi / 4 \quad ,
\end{equation}

where the simple relation between $A$ and $D$ applies for
off-lattice models with strictly circular cross section of the
cylinder (in the present Ising model case, $A=N_c (R)$, the number
of spins in a cross sectional plane for the considered choice of
$R$, i.e.~$N_c (3) =29$, $N_c(4)=49$ and $N_c(5)=69$). Now
$\sigma$ is the interfacial free energy per spin for the
three-dimensional Ising model. However, using the same reasoning
as in Eqs.~(\ref{eq8}),~(\ref{eq21}) we now obtain that the
effective transition temperature $T_0(L,D)$ of a long cylinder
from a multi-domain configuration to the single-domain
configuration is given by

\begin{equation} \label{eq33}
k_BT _0 (L,D) / \sigma (T_0)=A/\ln (L/3) \quad , \,\, L \, {\rm
large} \quad.
\end{equation}

While for $T \rightarrow 0$ again $\sigma (T) \rightarrow 2 J$ for
planar interfaces, for the temperatures of interest for the
present study Eq.~(\ref{eq33}) is not expected to be
quantitatively accurate (at not so low temperatures due to
boundary effects we expect that the actual interfacial energy
$F_{\rm int}$ is smaller than the asymptotic estimate $A \sigma$,
for the small radii $R$ studied here). Nevertheless,
Fig.~\ref{fig12} shows that $T_0(L,D)$ exhibits a distinct
decrease with increasing $L$, as expected.

\subsection{Ising cylinders with surface fields}

When the Ising model is re-interpreted as a lattice gas, it is
natural to assume that a ``free surface'' boundary condition
 is physically caused by a confining external wall, which then is
expected to provide an external potential to the particles
adjacent to the wall. In ``magnetic language'', such a wall
potential translates to a ``surface magnetic field'' $H_1$
\cite{107}.

While in the absence of $H_1$ the spin reversal symmetry of the
Ising model ensures that phase coexistence between domains of
opposite spontaneous magnetization occurs for bulk field $H=0$,
for $H_1 \neq 0$ this symmetry is broken. Thus, the lattice gas
model in thin film geometry with $H_1 \neq 0$ has been thoroughly
discussed as a model for capillary condensation
\cite{16,25,26,107,108,109}.

\begin{figure}
\includegraphics[scale=0.6]{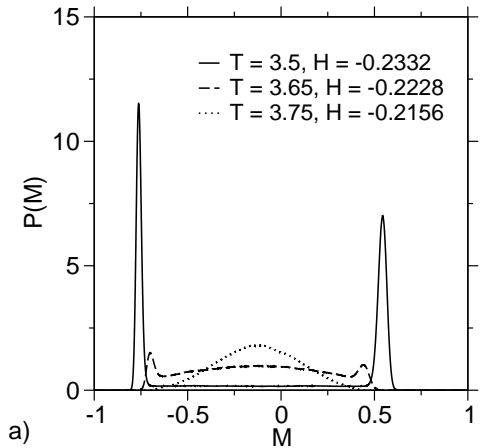}
\includegraphics[scale=0.6]{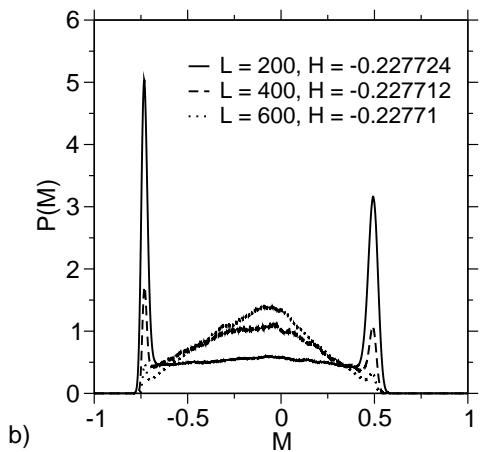}
\caption{a) Plot of $P_{L,D}(M)$ vs. $M$ for a cylinder of
radius $R=4$ and length $L=200$ for $H_1=0.75$ and three
temperatures $T=3.5, 3.65$ and $3.75$, as indicated. The chosen
field $H=H_{\rm coex}(T)$ in the bulk is also shown in the
figure. b) Same as a), but for $T=3.58$, $H=-0.2279$ (which was a
first rough estimate for $H_{\rm coex}(T))$ and three choices of
$L$, as indicated. The final estimates for $H_{\rm coex}(T)$ are
close to $H=-0.2277$ and were found by histogram extrapolation
methods, requiring an equal area rule for the two outer peaks.
Note that $P_{L,D} (M)$ no longer exhibits any symmetry with
respect to a sign change of $M$.} \label{fig13}
\end{figure}
Choosing a surface field $H_1=0.75$, $J=1$ that acts on all spins
in the surface layer (i.e., spins that have ``missing neighbors'')
we have to carry out scans where $H$ is varied to locate $H_{\rm
coex}(T)$, i.e.~the field where in short pores phase
coexistence occurs. At low temperatures, where for the considered
choices of pore length $L$ at phase coexistence only two peaks
occur in the distribution function $P_{L,D} (M)$
(cf.~Fig.~\ref{fig13}) an accurate sampling of $P_{L,D}(M)$ is
possible combining the standard Metropolis algorithm \cite{89}
with successive umbrella sampling \cite{110} methods. Note that
the single cluster algorithm \cite{92} or the related
Swendsen-Wang \cite{111} algorithm can be extended to include bulk
and surface fields but become very inefficient (apart from the
case where both surface and bulk fields become extremely small
\cite{109}) and then would not present any advantage.

The location of the coexistence field $H_{\rm coex}(T)$ in the
regime where $P_{L,D}(M)$ shows only two peaks can be based on the
``equal area rule'' \cite{112,113} as for phase transitions in the
bulk. Of course, in the present case, i.e.~for finite pore
diameter $D$, we still expect that this transition never becomes
sharp, irrespectively how large $L$ becomes. Thus, in the limit $L
\rightarrow \infty$ the susceptibility $\chi$, which now needs to
be defined by

\begin{equation} \label{eq34}
\chi= \beta N (\langle M^2 \rangle - \langle M \rangle ^2 )
\end{equation}

becomes a delta function (the limiting behavior for first order
transitions \cite{62,112}) only for $T \rightarrow 0$. In the case
of $L$ finite where at low $T$ only two peaks in $P_{L,D} (M)$
occur, we expect that $\langle M^2  \rangle - \langle M \rangle^2
$ reaches at $H=H_{\rm coex}(T)$ a maximum of order unity which we
denote as $c_{\rm max}(T)$ (since the spin reversal symmetry is
broken due to the surface field, we can no longer conclude that
$\langle M \rangle =0$ at $H=H_{\rm coex}(T))$. Thus we conclude
that $\chi$ reaches a maximum value

\begin{equation} \label{eq35}
\chi_{\rm max}(T)= \beta N c_{\rm max} (T), \quad H=H_{\rm coex}
(T)
\end{equation}

while in the region of fields where $P_{L,D}(M)$ has a single
maximum only (at the considered low temperature where $\xi_D(T)
\gg L)$ the susceptibility $\chi$ will be of order unity.
Following the reasoning of \cite{113} we can conclude that in this
region the rounding of the transitions is simply given by the
condition that

\begin{equation} \label{eq36}
\chi_{\rm max} \Delta H \leq 1 \quad , \quad \beta \Delta H
\propto 1/N \quad , \Delta H \equiv H-H_{\rm coex} (T) \quad .
\end{equation}

Since $N$ is very large throughout our study, $\Delta H$ is
very small, and $H_{\rm coex}(T)$ in this regime of low
temperatures is rather well defined.

Of course, the situation becomes more subtle at higher
temperatures, where the third peak in the distribution
$P_{L,D}(M)$ appears (Fig.~\ref{fig13}). As long as the weight of
this central peak is not yet much larger than the weight of the
two other, sharper, peaks, we simply can ignore this peak and
still apply the equal weight rule with respect to the two outer
peaks. However, the equal weight rule method for estimating
$H_{\rm coex}(T)$ becomes obsolete when the weights of the two
outer peaks become relatively small (and ultimately the two outer
peaks completely disappear!)

Thus, we resort to a general alternative method to estimate
$H_{\rm coex}(T)$, which requires to scan $\chi(H)$, as defined in
Eq.~(\ref{eq34}), as a function of the field $H$. This method
would have been very inconvenient at low temperatures - the
increments $\delta H$ of the field $H$ necessary to perform such a
scan would need to satisfy the condition $\delta H \ll \Delta H$
and since $\Delta H$ is so small \{Eq.~(\ref{eq36})\} and $H_{\rm
coex} (T)$ is not known beforehand, an enormous (and not reasonable)
effort of computer resources would be implied (and furthermore the
transition  would easily be missed due to hysteresis). However, for
$T \geq T_0(L,D)$ hysteresis is no longer a severe problem, and
the rounding of the transition is much smaller, since now ($N_c$
is the number of spins in a cross sectional plane, $N=N_cL$)

\begin{equation} \label{eq37}
\chi_{\rm max} (T) \approx 2 \beta \xi _D(T) N_c \, \, c_{\rm max}
(T)
\end{equation}

and thus $\chi_{\rm max}(T)$ is much smaller than in the region
where Eq.~(\ref{eq34}) holds. So $\Delta H$ is no longer so small;
furthermore, one can get a first estimate for $H_{\rm coex}(T)$ by
an extrapolation from the region where $P_{L,D}(M)$ has only two
peaks, and the analysis as described above works.

\begin{figure}
\includegraphics[scale=0.6]{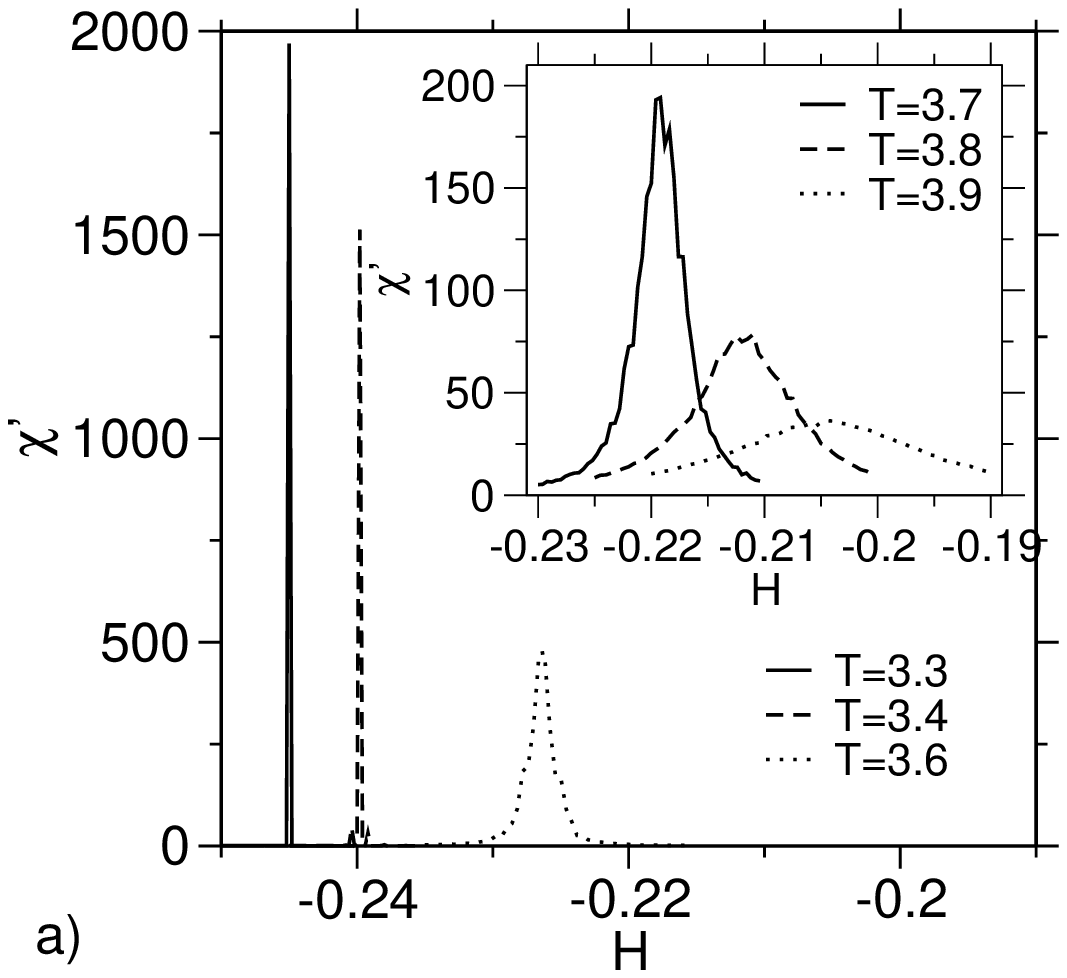}
\includegraphics[scale=0.6]{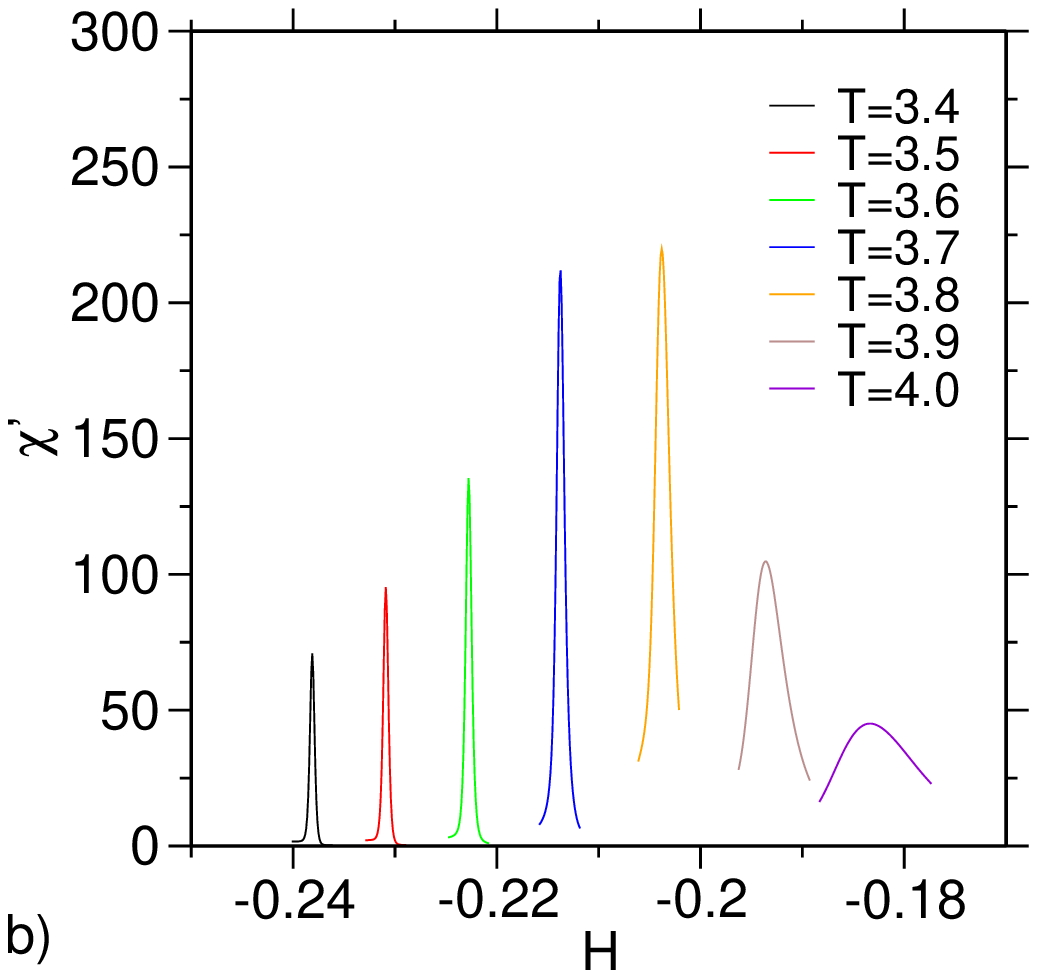}
\caption{Plot of $\chi'$ vs. $H$ for the case $H_1=0.75$,
$R=4$, $L=300$ and increasing temperatures from left to right obtained by explicit simulations at each value of $H$. b) Plot of
$\chi'$ vs. $H$ for the case $H_1=0.75$, $R=5$, $L=200$ and
several temperatures as indicated using extrapolation to different values of $H$.} \label{fig14}
\end{figure}
Thus, we have scanned the region of interest choosing steps
$\Delta H =0.0002$, carrying out runs with $2$ million MC steps per spin at each
state point ($H,T$). Histogram reweighting methods were used to
improve the accuracy of the results, as is standard \cite{89}.
Fig.~\ref{fig14}a presents typical data for the case $R=4$,
$L=300$. Since we have found that also $\chi'$ as defined in
Eq.~(\ref{eq31}) has a sharp peak at $H \approx H_{\rm coex}$ we
used the location $\chi'_{\rm max}$ as well. Note that in the
region where $P_{L,D}(M)$ has a single peak, the width of this
peak is rather narrow if $H$ differs appreciably from $H_{\rm
coex}(T)$, since there then the state of the pore is uniform, no
nucleation of two-phase fluctuations takes place. Then $\langle M
^2 \rangle \approx \langle |M| \rangle ^2$ irrespective of the
value of the peak, and $\chi'$ is of order unity. Only for $H$
near $H_{\rm coex}$ will the distribution $P_{L,D}(M)$ show some
anomalous broadening, resulting from the fluctuations associated
with the coexistence of multiple domains. Therefore, recording the
maxima of $\chi'$, which at temperatures near $T_0(L,D)$ are much
easier to sample (Fig.~\ref{fig14}b), is a useful method to
estimate $H_{\rm coex}(T)$ for $T \geq T_0(L,D)$, see
Fig.~\ref{fig14}b.

\begin{figure} 
\includegraphics[scale=0.28]{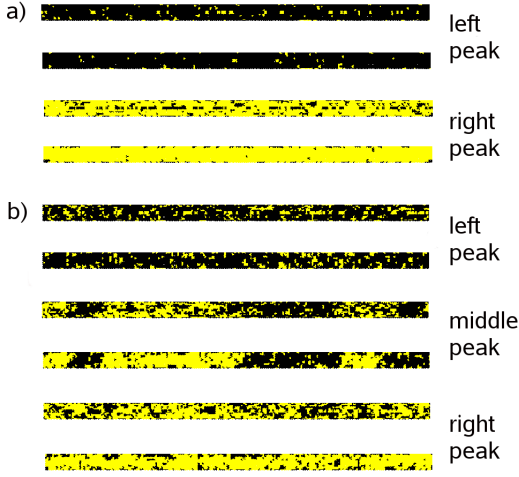}
\caption{Configuration snapshots of a system with $H_1=0.75$,
$R=4$, $L=200$ at $H_{\rm coex} (T)$ for $T=2.5$ (a) and $T=3.65$
(b). Each snapshot shows for each peak of $P_{L,D}(M)$ an outside
projection (upper part) and a cross section containing the
cylinder axis (lower part). Negative spins are shown in black,
positive spins in yellow. Note that for $T=2.5$ one has two
peaks for $P_{L,D} (M,T)$, while for $T=3.65$ one has three peaks.
The snapshots for the middle peak in part (b) show the multiple
domain structure clearly.}\label{fig15}
\end{figure}
\begin{figure}
\includegraphics[scale=0.6]{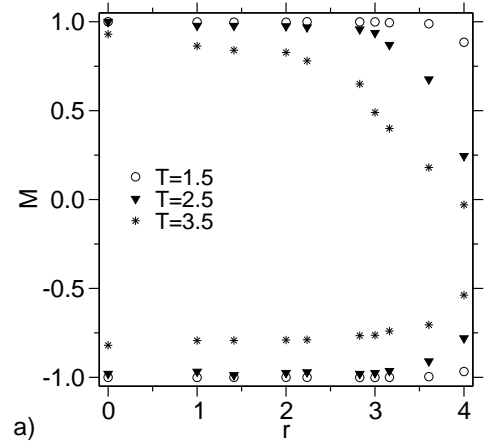}
\includegraphics[scale=0.6]{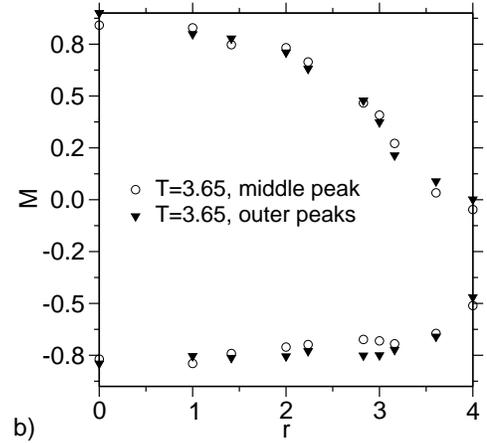}
\caption{a) Magnetization profiles $m(r)$ as a function of the distance $r$
from the cylinder axis, for $H_1=0.75$, $R=4$, $L=200$, and
$T=1.5$, $2.5$ and $3.5$, as indicated. The upper set of curves
belongs to states with positive magnetization of the pore, the
lower set for states with negative magnetization, for $H=H_{\rm
coex}(T)$. In all cases $P_{L,D}(M)$ still has only two peaks. b)
Magnetization profiles $m(r)$ as a function of the distance $r$
for $T=3.65$, showing both profiles from the peaks representing
the ``pure phases'' and from the right and left parts of the
middle peak. These latter profiles were extracted from local
slices through the cylinder (the contribution of slices containing
the interfaces turn out to be still negligible at this
temperature.)} \label{fig16}
\end{figure}
It is interesting to note that the coexisting phases of the
cylinder (in the region $T\leq T_0(L,D))$ are inhomogeneous. This
is evidenced both by snapshot pictures (Fig.~\ref{fig15}) and
radial order parameter profiles (Fig.~\ref{fig16}) taken for our
systems. One can see that on the outside surface of the cylinder
(seen in the projection snapshots of Fig.~\ref{fig15}) there is
always more disorder. The reduction of the local magnetization at
the surface, when the bulk of the cylinder has positive
magnetization, can be interpreted as a precursor of wetting
phenomena. Of course, true wetting layers cannot form in nanopore
cylinders, and hence we also do not find a transition as proposed
by Liu et al. \cite{67}. Crossing the wetting transition
temperature $T_w(H_1)$ \cite{16,37,38,39,40} it was predicted that
a transition from ``plugs'' to ``capsules'' should occur
\cite{67}, and an attempt was made \cite{114} to locate this
transition by Monte Carlo simulations in $L_\bot \times L_\bot
\times L$ systems with $L_\bot =14,20$ and $28$, varying $L$ from
$L=40$ to $L=320$, and various choices of $H_1$. However, for such
rather wide and short pores the problem of multiple domains did
not yet arise, and the issues about intrinsic rounding of
transitions in the quasi-one-dimensional pore geometry were not
studied in these investigations \cite{67,114}.

\begin{figure}
\includegraphics[scale=0.6]{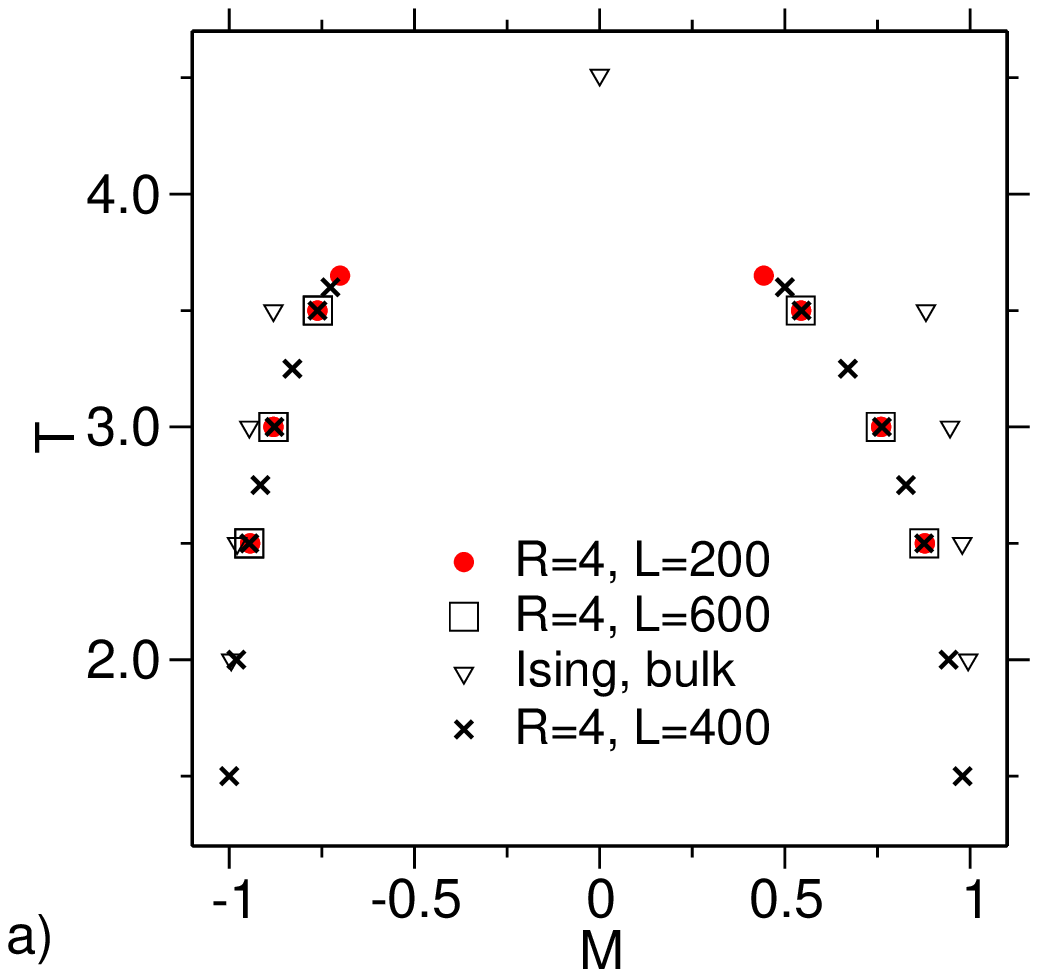}
\includegraphics[scale=0.6]{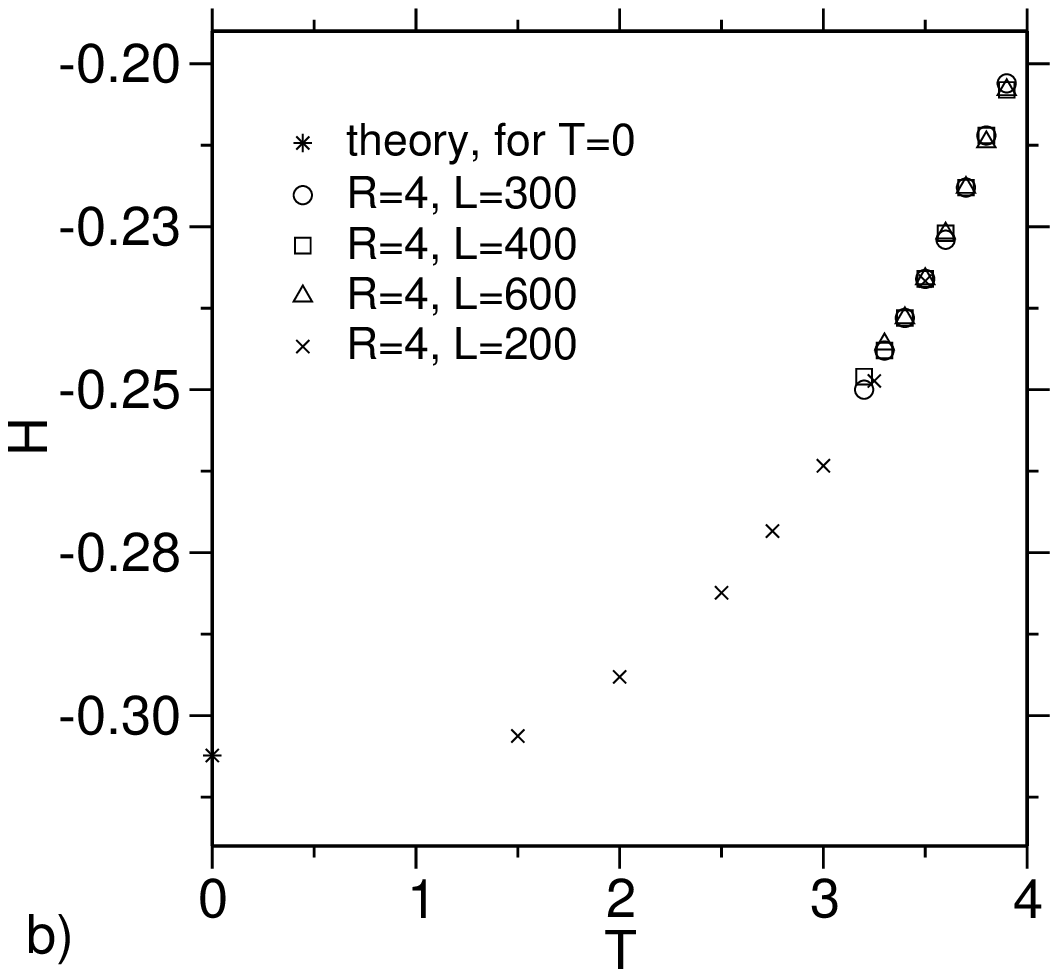}
\caption{``Phase diagram'' of the cylindrical Ising pore
for the case $R=4$, $H_1=0.75$, and three choices of $L$ plotted
in the (T,M) plane (a) and in the (H,T) plane (b). In part (a),
data for $M_{\rm coex} (T)$ for the spontaneous magnetization of a
three dimensional bulk Ising model are included. The data for the
two coexisting phases in (short) cylinders are extracted from the
positions of the two peaks at $H=H_{\rm coex}(T)$,
respectively.}\label{fig17}
\end{figure}
\begin{figure}
\includegraphics[scale=0.6]{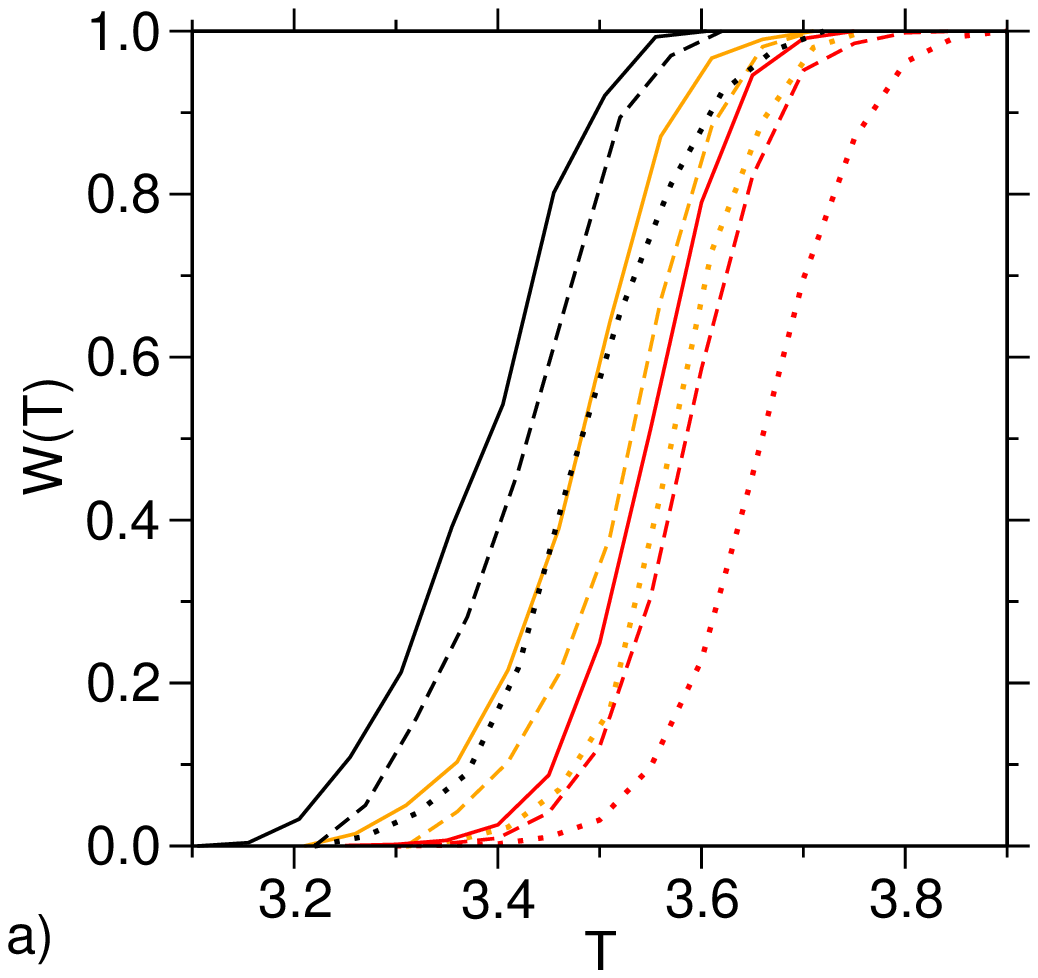}
\includegraphics[scale=0.6]{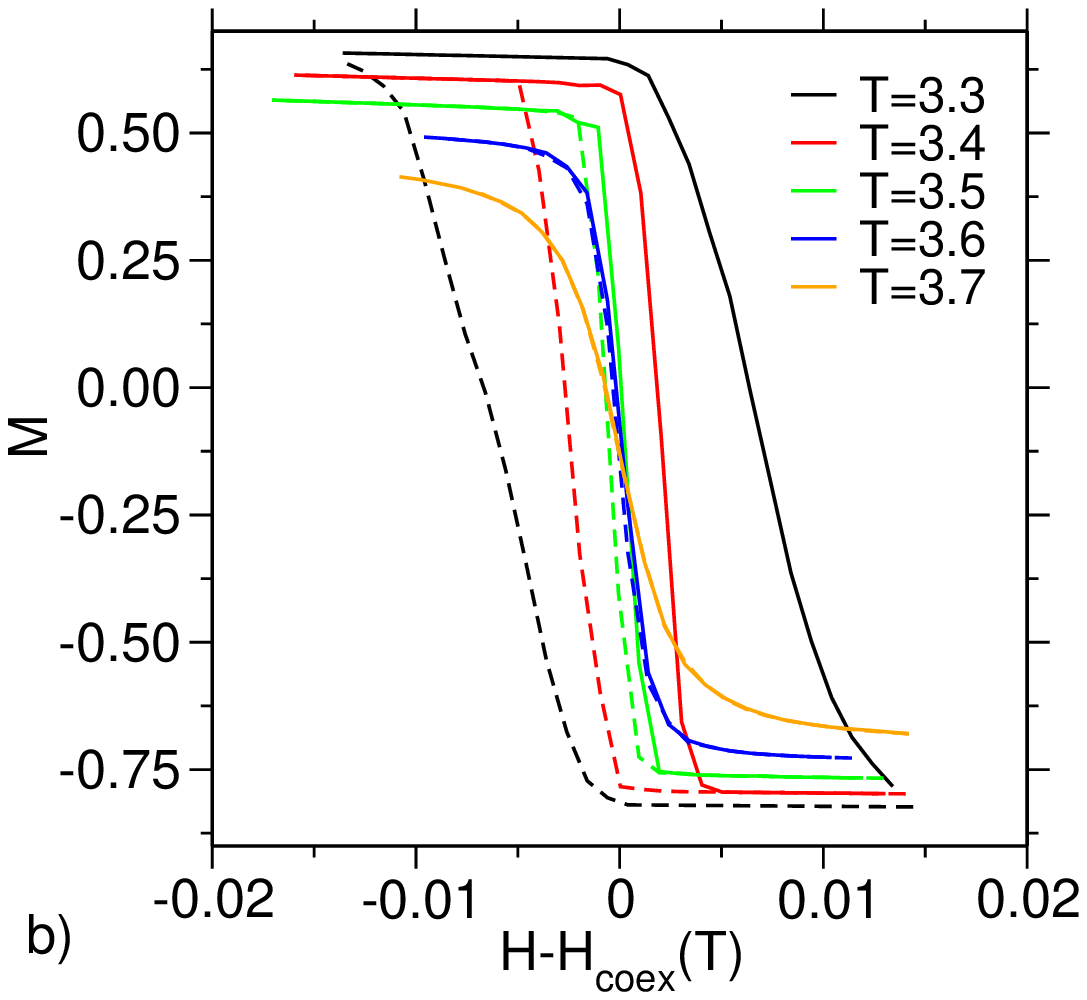}
\caption{a) Weight $W$ of the central peak of
$P_{L,D}(M)$ plotted vs. temperature for Ising pores of radius
$R=4$ and three choices of the surface field, $H_1=1.5$, $H_1=0.75$
and $H_1=0$ (from left to the right in different colors). In each case the pore lengths
$L=600$, $400$, and $200$ are included (from the left to the right with different line styles).
b) Magnetization of Ising cylinders of radius $R=4$, length $L=200$,
surface magnetic field $H_1=0.75$, plotted vs. field $H-H_{coex}(T)$
at $5$ temperatures, as indicated. Bulk field $H$ was varied in steps of $0.001$.
}\label{fig18}
\end{figure}
As a final point of this section, we present in Fig.~\ref{fig17}
the ``phase diagram'' of our model, both in the $T-M$ plane and the
$H-T$ plane. Note that this ``phase diagram'' is only meant to
describe the phase coexistence that persists if $\xi_D(T) \gg D$
on a local scale in long cylinders (or in short cylinders, if
$\xi_D(T)$ still exceeds $L$). On the scale of the axes chosen in
Fig.~\ref{fig17}, the ``phase boundaries'' still look sharp,
although the transition line in Fig.~\ref{fig17}b is
intrinsically rounded over some width $\Delta H$, but for the
temperature region shown, the rounding is still small. However,
this phase diagram cannot uniquely be continued up to a ``capillary
critical point'': when $\xi _D(T)$ has decreased to a value
comparable to $D$, the rounding gets very strong, and different
criteria to locate a ``transition'' will no longer coincide (e.g.,
the position of a maximum for $\chi(H)$ and $\chi'(H)$ will no longer
 agree, etc.) Thus, our ``phase diagram'' ends in an ``open
way'' at $T=4.0$: for pores as narrow as $R=4$, the difference
between the order parameter of the two coexisting phases then has
already decreased significantly, and at slightly higher
temperatures it is no longer possible to distinguish the regions
of the ``pure coexisting phases'' inside the pore from the
interfaces separating them. However, it is always of interest to
study in very long pores the transition at $\xi_D(T) \approx L/3$
from the multiple domain phase coexistence at $H_{\rm coex}(T)$ to
the ``pure'' coexisting phases inside the pore. Fig.~\ref{fig18}a
compares this transition for three choices of $H_1$: we see that
increasing the surface field shifts the transition to lower
temperatures, but the qualitative characteristics stay the same.

As in the case of the Ising strip, we can verify that in the same region of
temperatures where the change of $P_{L,D}(M)$ from the double-peak
distribution to the triple-peak distribution occurs (cf. Fig.~\ref{fig13}a)
the hysteresis in the magnetization process vanishes (Fig.~\ref{fig18}b),
namely near $T\approx 3.6$. The figure shows that at this temperature one still
can identify clearly the difference in order parameter of the
vapor-like and liquid-like branch of the lattice gas model.
The data in Fig.~\ref{fig18}b) are for a rather short pore, namely
$L=200$: It is clear (cf. also Fig.~\ref{fig13}b) that for a 
longer $L$ this change of $P_{L,D}(M)$ occurs for lower
temperatures, and also the onset of hysteresis occurs at the
lower temperature the larger the pore length $L$ is considered.

\section{Colloid-Polymer Mixtures Confined in Cylinders: A Monte
Carlo Study of the Asakura-Oosawa Model}

Colloidal dispersions have become model systems for the study of
phase behavior of condensed matter, since the large size of the
colloidal particles allows the use of experimental observation
techniques that cannot be used for small molecular systems.
Furthermore, interactions among colloidal particles are tunable
to a large extent \cite{115,116,117,118,119}. Colloid-polymer
mixtures \cite{120,121,122,123,124} have been particularly
suitable to study liquid-vapor-like phase separation into
colloid-rich and colloid-poor phases, including their interfacial
behavior. There also exists a very simple theoretical model, due
to  Asakura and Oosawa \cite{70} and Vrij \cite{125} (henceforth
referred to as ``AO model''), well suited for Monte Carlo
simulation studies \cite{34,126,127,128,129,130,131,132}. In this
model, colloids are simply described as hard spheres of radius
$R_c$ while polymers are soft spheres of radius $R_p$. While
overlap among colloids and between polymers and colloids is
strictly forbidden, i.e. the potential energy is given by

\begin{equation} \label{eq38}
U_{cc}(r < 2 R_c) = \infty, \quad U_{cc} (r \geq 2 R_c=0) \quad ,
\end{equation}

\begin{equation} \label{eq39}
U_{pc} (r<R_p + R_c) = \infty, \quad U_{pc} (r \geq R_p + R_c)=0
\quad,
\end{equation}

two polymer coils can interpenetrate and hence overlap with no
energy cost, $U_{pp}(r)=0$ irrespective of distance $r$. Vink et
al. \cite{130,131,132} have already performed an extensive study
of capillary condensation for colloid-polymer mixtures confined
between two parallel hard walls a distance $D$ apart, and have
shown that for distance $D$ of the order of a few colloid
diameters a crossover from three-dimensional to two-dimensional
Ising critical behavior occurs. Due to the large sizes of colloid
particles, it should be experimentally feasible to also study
capillaries which are only a few colloids' diameters wide, and since
for particles in the size range of a $\mu m$ the atomistic
corrugation of real walls clearly is negligible, fairly ideal
conditions should be realizable.

In the present work, we have extended this work
\cite{127,128,129,130,131,132} to confinement in cylindrical pores
of diameters $D= 12 R_c$ and lengths $L$ up to $L=540 R_c$, for
$R_p/R_c=0.8$. In the following, $R_c=1$ shall be used as unit of
length in this section. The Monte Carlo simulations were carried
out in the grand-canonical ensemble, choosing the chemical
potential $\mu_c$ of the colloids and the polymer reservoir
packing fraction $\eta^r_p$

\begin{equation} \label{eq40}
\eta^r_p=(4 \pi R^3_p/3) \exp (\mu_p/k_BT) \quad,
\end{equation}

\begin{figure}
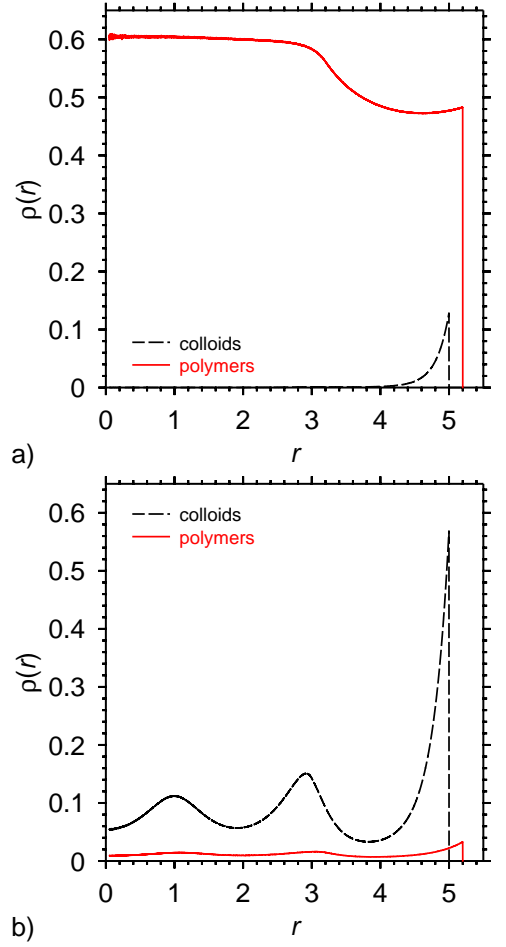

\includegraphics[scale=0.5]{Fig_19a.ps}
\includegraphics[scale=0.5]{Fig_19b.ps}
\caption{Radial density of colloids ($\rho_c(r))$ and polymers
$(\rho_p(r))$ plotted vs. distance from the cylinder axis for
$\eta_p^r =1.30$, $L=60$, $D=12$. Case a) shows these profiles in
the vapor-like phase, case (b) for the liquid-like
phase.}\label{fig19}
\end{figure}
\begin{figure}
\includegraphics[scale=0.45]{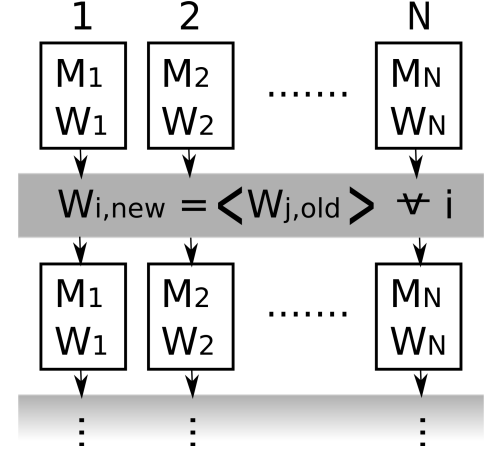}
\caption{Parallelization scheme of the ``Wang-Landau`` algorithm. The first row of
numbers is the CPU index. $M_i$ is the number of Monte Carlo steps, which is performed.
$W_i$ is the weight function of CPU $i$. The brackets denote an average weighted by the 
number of MC steps. The average replaces the weight function $W_{i}$ of every process $i$.}
\label{fig20_WL}
\end{figure}
where $\mu_p$ is the chemical potential of the polymers, as
independent control variables. Observables of interest then are
both global average densities $\rho_p=N_p/V$,
$\rho_c=N_c/V$ of colloids and polymers ($N_p$, $N_c$ are
the total number of polymers and colloids in the volume $V=(\pi
D^2 / 4)L$ as well as the corresponding radial density profiles
$\rho_p(r)$, $\rho_c(r)$, see Fig.~\ref{fig19}. One can see that
in the  vapor-like phase the polymer density is reduced near the
pore wall, while colloids are attracted to the pore wall both in
the vapor-like and liquid-like phase. Note that phase coexistence
in the pore was located as for the thin film case by scanning the
chemical potential $\mu_c$ until one finds a double-peak
distribution, where then the equal weight rule \cite{112,113} is
applied to estimate the value of the chemical potential at
coexistence $\mu_{c,\rm coex}$. As for the case of the AO model in
the bulk \cite{127,128,129} and in thin film geometry
\cite{130,131,132}, cluster moves \cite{127} and successive
umbrella sampling methods \cite{110} are applied throughout. 
For large systems a parallel version of the ``Wang-Landau'' algorithm \cite{133} 
was implemented. The idea, schematically shown in Fig.~\ref{fig20_WL}, 
is to correlate a priori independent simulations by taking averages 
over the weight functions iteratively generated by the ``Wang-Landau'' 
algorithm on each CPU. The average between the weight functions of the 
single simulations is weighted with the MC steps done so far and is used 
as a new weight function for all CPUs. This procedure uses only a very 
small amount of communication and scales therefore almost linearly up to $1440$ CPUs. 
In comparison to the same number of non-communicating independent simulations, 
a reduction of the systematic error of the biasing algorithm was observed, 
which leads to a faster convergence of the weight function to the true free energy landscape.

\begin{figure}
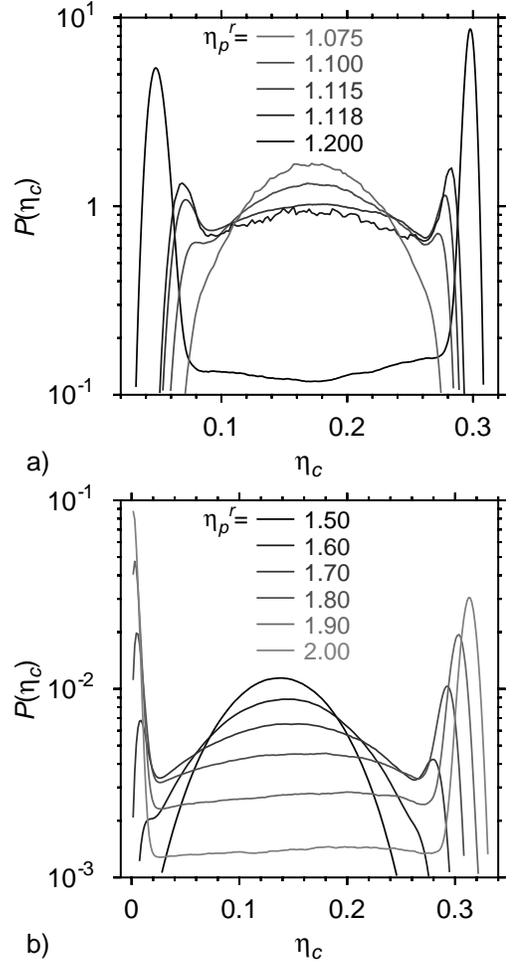

\includegraphics[scale=0.5]{Fig_21a.ps}
\includegraphics[scale=0.5]{Fig_21b.ps}
\caption{Probability distribution $P_{L,D}(\eta_c)$ of the colloid packing
fraction in a cylinder of diameter $D=12$ and length $L=180$ (a) 
and for $D=6$, $L=100$ (b).}\label{fig21}
\end{figure}
\begin{figure}
\includegraphics[scale=0.11]{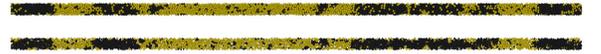}
\caption{Snapshots for the system with $D=12$ and $L=540$ at the
polymer reservoir packing fraction $\eta_p^r=1.15$. Colloids are
shown in yellow. The lower picture shows the cylindrical simulation
box sliced at the plane $x=0$, while the upper picture visualizes
the projection of the particles at the confining border.}
\label{Snap}
\end{figure}
While for the case shown in Fig.~\ref{fig19} the state of the
cylinder at phase coexistence is axially homogeneous, and this
fact also shows up in the probability distribution $P_{L,D}
(\eta_c)$, $\eta_c=(4 \pi R^3_c/3) \rho_c$ being the colloid
packing fraction, Fig.~\ref{fig21},  since $P_{L,D}(\eta_c)$
just has two peaks and is flat in between, at lower values of
$\eta^r_p$ one again finds a distribution with three peaks.
 As in the Ising case, the interpretation of
the distribution exhibiting a ``central'' peak is the formation of
multiple domain walls across the pore (Fig.~\ref{Snap}).

\begin{figure}
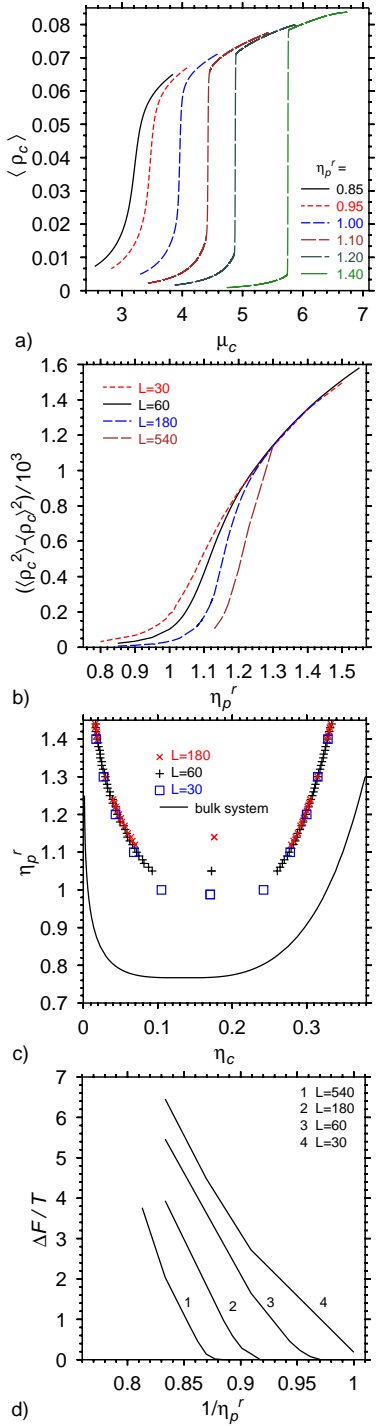

\includegraphics[scale=0.375]{Fig_22a.ps}\ \ \ \ \ \ 
\includegraphics[scale=0.375]{Fig_22b.ps}\\
\includegraphics[scale=0.375]{Fig_22c.ps}\ \ \ \ \ \ 
\includegraphics[scale=0.375]{Fig_22d.ps}
\caption{a) Average density $\langle \rho_c \rangle$ of the
colloids in a pore of linear dimensions $L=60$, $D=12$ plotted vs.
$\mu_c$ at various values of $\eta_p^r$ as indicated. b) Plot of
the maximum value of the density fluctuation for $D=12$ and
various $L$ as indicated, plotted vs. $\eta^r_p$. c) Phase diagram
of the AO model in the plane of variables ($\eta_p^r$, $\eta_c)$
shown for cylinders of diameter $D=12$ and three choices of $L$.
Full curve shows the coexistence curve for the corresponding bulk
AO model. The symbols at $\eta _c \approx 0.16$ to $0.17$ show the
transition from the single-domain to the multiple domain state in
the pore. d) Barrier $\Delta F/T$ against nucleation of interfaces 
in the AO model confined to a cylindrical pore of diameter $D=12$
plotted versus inverse polymer reservoir packing fraction.} \label{fig22}
\end{figure}
Fig.~\ref{fig22}a shows the average colloid density $\langle\rho_c\rangle$
as a function of $\mu_c$. The maximum value of the fluctuation
$\langle (\rho_c -\langle \rho_c\rangle)^2 \rangle$ studied as a
function of $\eta^r_p$ for several choices of $L$ is shown in Fig.~\ref{fig22}b. Also the
corresponding phase diagram is shown (Fig.~\ref{fig22}c). While in
the bulk well-defined vapor-liquid like phase coexistence occurs,
ending in a critical point at $\eta^r_{p,cr}=0.765$,
$\eta_{c,cr}=0.13$, the phase coexistence in the cylindrical pore
exists over a finite correlation length $\xi_D$ only. The value of
$\langle\eta_c\rangle$ in the coexisting vapor-like and liquid-like phases
depend on $L$ only very weakly: however, the larger $L$ the larger
$\eta_p^r$ has to be chosen to ensure that one still has phase
coexistence between single-domain states in the pore, rather than
a multiple domain structure.

\begin{figure}
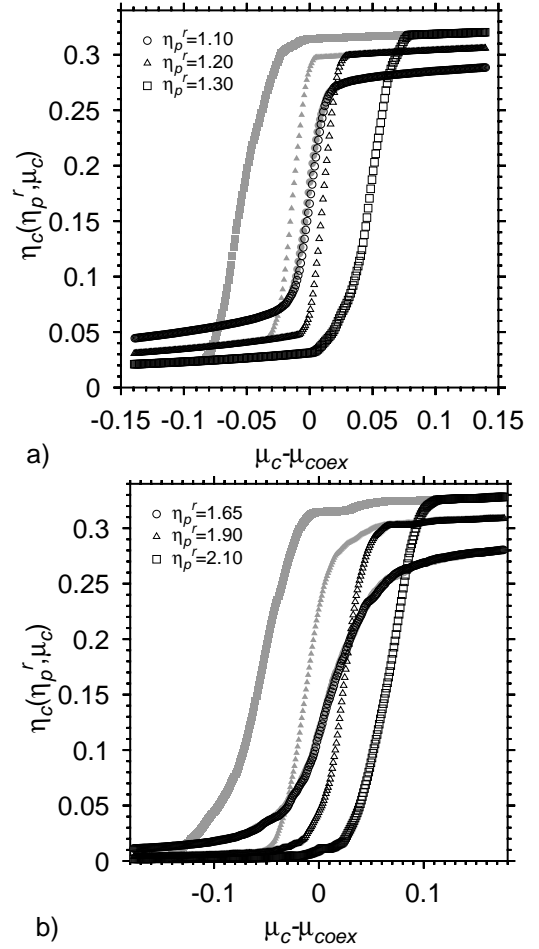

\includegraphics[scale=0.5]{Fig23a_Av.ps}
\includegraphics[scale=0.5]{Fig23b_Av.ps}
\caption{Two hysteresis plots for the AO model. The chemical potential was
varied in steps of $0.001 k_BT$. Several simulation runs (up to 38) were
averaged. For high polymer reservoir packing fractions large sample to sample
fluctuations occur. The open symbols show data for which the chemical
potential was increased step-wise while the full symbols show data for
which the value of the chemical potential was decreased step by step.
(a) shows the disappearance of the hysteresis for a system with $D=12$
and $L=180$. (b) shows the disappearance of the hysteresis for a system with $D=6$
and $L=100$.} \label{fig24}
\end{figure}
Again it is of central importance to verify the connection between
the change of the distribution $P(\eta_c)$ with decreasing
$\eta_p^r$ from the double peak behavior at large $\eta_p^r$
to the three-peak behavior at somewhat smaller $\eta_p^r$
(cf.~Fig.~\ref{fig21}) and the disappearance of hysteresis
at a value of $\eta_p^r$ which is still distinctly larger
than the pore critical temperature (where in Fig.~\ref{fig22}c
the vapor-like and liquid-like branches of the coexistence curve
of the fluid confined in the pore have merged). This
connection is verified by Fig.~\ref{fig24}, which shows that
for $D=12$ and $L=180$ hysteresis indeed disappears in
between $\eta_p^r=1.2$ and $\eta_p^r=1.1$, while the
coexistence curve branches exist up to about $\eta_p^r=1.0$
(Fig.~\ref{fig22}c). Again we predict that this difference between
the pore critical temperature and the hysteresis critical
temperature should increase with $L$. As an analogue to Fig.~\ref{fig8}
the free energy barriers are shown in Fig.~\ref{fig22}d for various choices
of $L$.

\begin{figure}
\includegraphics[scale=0.4]{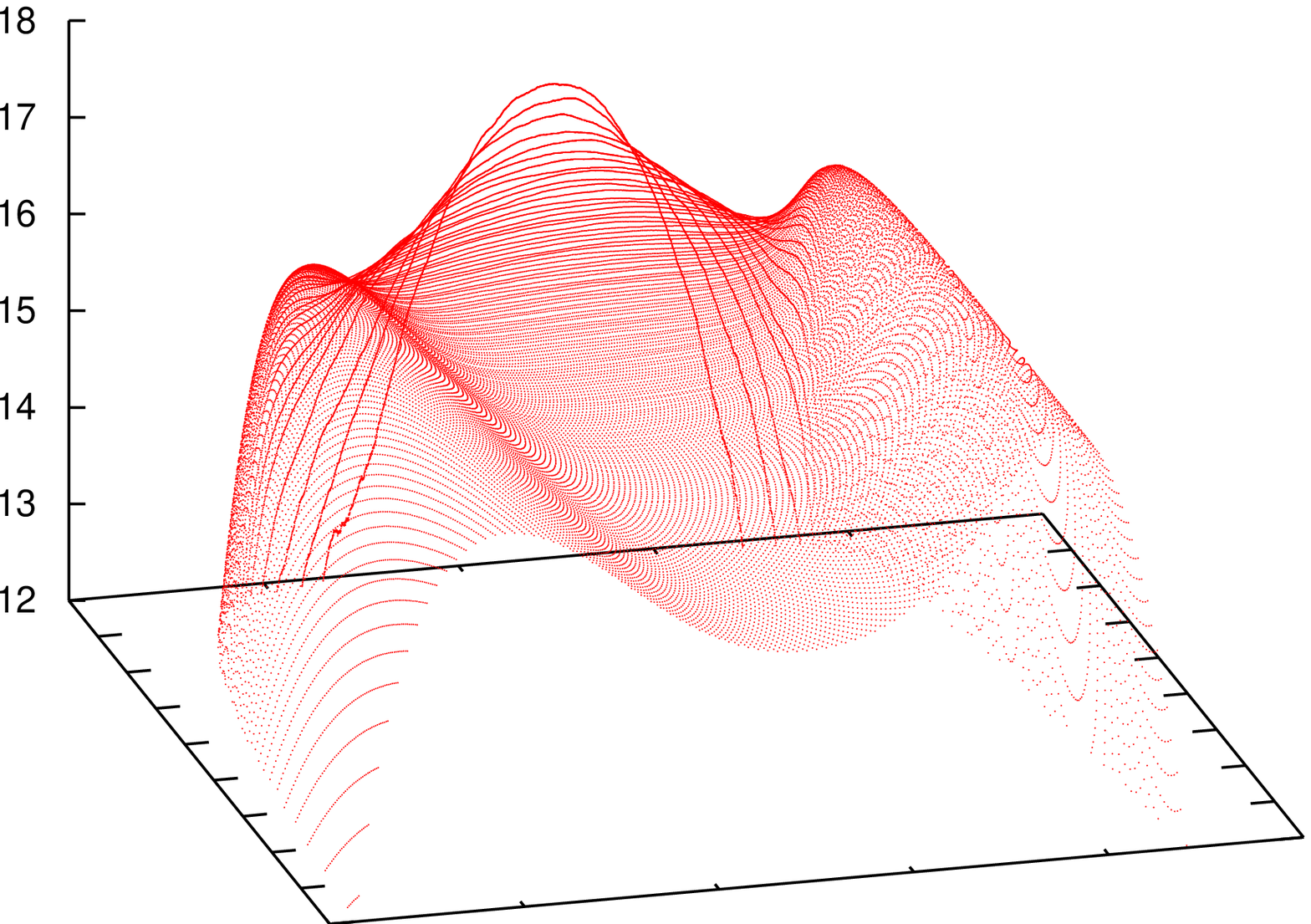}
\includegraphics[scale=0.45]{Fig_25a_2.ps}
\includegraphics[scale=0.4]{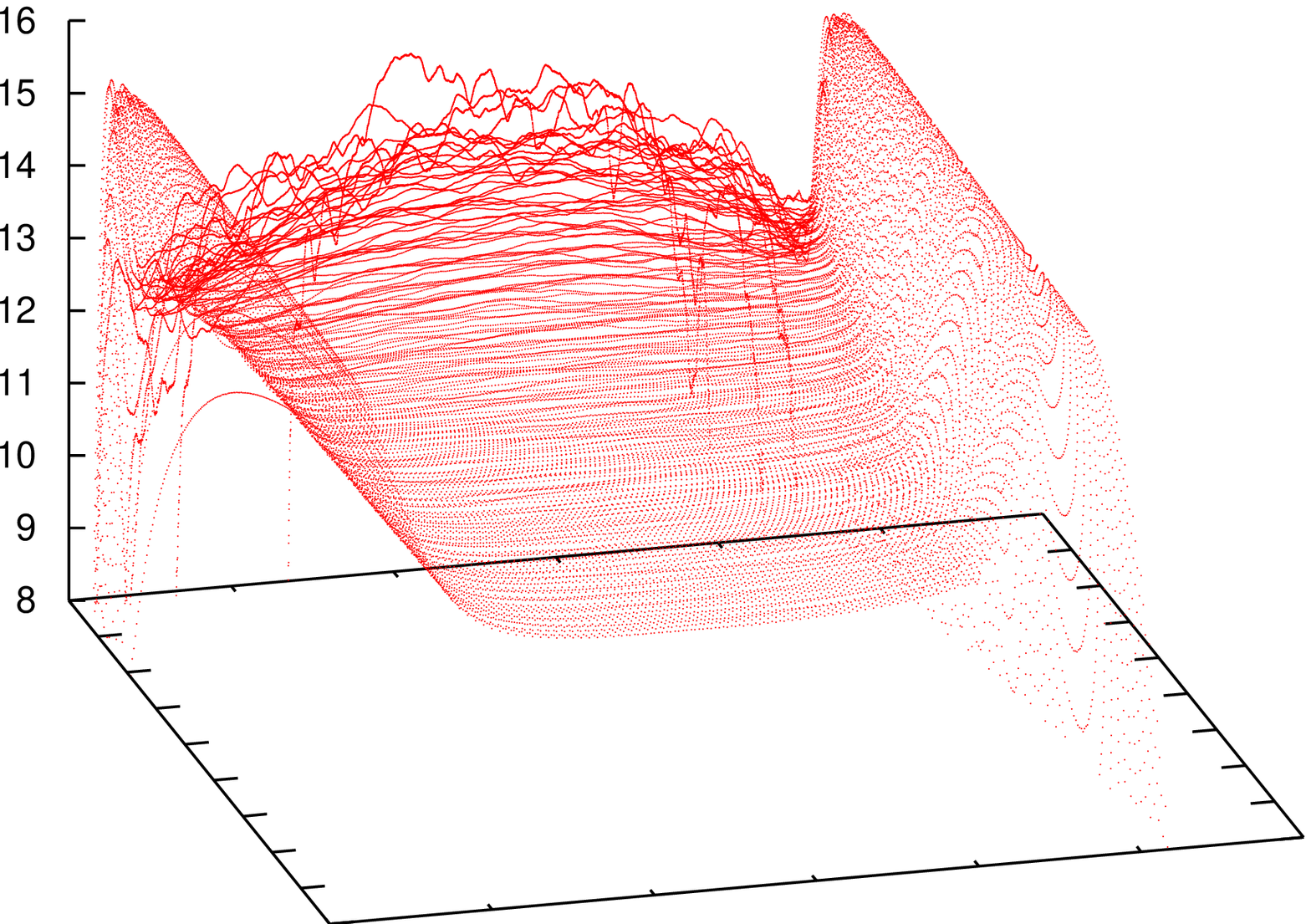}
\includegraphics[scale=0.45]{Fig_25b_2.ps}
\caption{Three-dimensional plot (left part) and contour
plot (right part) of $P_{D,N_s}(N_c)$ vs. number of subsystems
$N_s$, for $\eta^r_p=1.1$ (first two graphs) and $\eta_p^r=1.5$ (last two graphs). Note that
the number of colloids $N_c$ is normalized by the length $L/N_s$.} \label{fig23}
\end{figure}
An alternative way to explore this transition from axially
symmetric phase coexistence in the cylindrical pore to a multiple
domain structure uses a very long pore ($L=1800$) which is cut into
a one-dimensional array of $N_s$ subsystems, and recording the
distribution $P_{D,N_s} (N_c)$ of the number of colloids in the
subsystems (Fig.~\ref{fig23}). One can nicely see that for short
enough subsystems (i.e., for $N_s \geq 60$) the subsystem is typically homogeneous, since
there occur just two peaks with a minimum in between. However, for
very large $L/N_s$ one still finds the middle peak, as a signature
of the multiple domain structure, and the transition between both
types of behaviors (as a function of $L/N_s$ or $\eta^r_p$,
respectively) is completely gradual. Thus, the fact that the
coexisting phases in the phase diagram of Fig.~\ref{fig22}c show
practically no $L$-dependence, and the fact that the equilibrium
isotherms (Fig.~\ref{fig22}a) at large $\eta^r_p$ have an almost
perpendicular part should not be taken as evidence that in the
cylindrical pore a sharp, well-defined phase transition exists: as
in the Ising model, the transition is rounded, but for large
$\mu^r_p$ the extent of rounding is very small.

\section{Conclusions}

In this paper, the characteristic features of phase transitions of
Ising-like systems in a quasi-one dimensional geometry have been
explored by Monte Carlo simulations for four generic models: (i)
Ising $L \times D$ strips with $L \gg  D$ and periodic boundary
conditions throughout (ii) Ising ``cylinders'' of length $L$ and
cross section containing $N_{cr}$ sites enclosed by a circle of
radius $R$, with a ``missing neighbor'' boundary condition that
does not destroy the symmetry between the coexisting phases in the
ground state; (iii) the same model as in (ii), but with a surface
field $H_1$ acting on the spins which have ``missing neighbors'',
so that the spin reversal symmetry is broken, and the model
(interpreted as a lattice gas) exhibits capillary condensation;
and (iv) the AO model confined to cylindrical pores of diameter
$D$ and length $L$, confined by hard repulsive walls, as an
off-lattice model that lacks any particular symmetries already in
the bulk.

We have shown that all models exhibit qualitatively similar
behavior, namely two strongly rounded transitions occur when at
phase coexistence conditions the temperature (or temperature-like
variable, such as ($n^r_p)^{-1}$ in the case of the
colloid-polymer mixture, respectively) is lowered: at a
temperature rather close to the critical temperature of the bulk,
a rounded transition occurs from the disordered phase (which is
axially symmetric but may have nontrivial order parameter profiles
in the phase perpendicular to the cylinder axis, induced by the
boundaries, if there is no complete symmetry between the
coexisting phases) to a locally ordered phase, where the size of
the domains $\xi_D(T)$ in axial direction exceeds distinctly the
pore diameter, so that a long cylinder ($L \gg \xi _D (T) \gg D)$
is characterized by a sequence of interfaces across the cylinder
axis. The order parameter distribution at coexistence is then a
very broad Gaussian characterized by a very large response
function (if the transition is studied as a function of the field
conjugate to the order parameter, the rounding is exponentially
small in the cross-sectional areas of the cylinder). At $L/3
\approx\xi_D(T)$, i.e.~at $T=T_0(L,D)$, a second, again rounded,
transition occurs, where the state of the system is again
axially uniform and either the vapor-like or liquid-like phase
dominates. When one studies the kinetics of the transition between
vapor-like and fluid-like phases, varying the field conjugate to
the order parameter, one finds pronounced hysteresis in this low
temperature region, $T < T_0(L,D)$ where $\xi_D(T) > L$, but these
hysteresis loops get narrow when $T \approx T_0(L,D)$ and vanish
completely for $T>T_0(L,D)$. Thus, we suggest that the
``hysteresis critical point'' can be associated with the lower
temperature $T_0(L,D)$ rather than the upper pseudo-critical
temperature of the capillary (where $\xi _D(T) \approx D$ and the
difference in order parameter between the coexisting phases
disappears).
A prediction that could be tested experimentally is our result
that the difference between this ``hysteresis critical temperature''
and the ``pore critical temperature'' should increase
logarithmically with the length $L$ of the cylindrical pore.
We hope that our study stimulates additional
experimental work using pores of both well-controlled diameter and
length to check our predictions and thus confirm that a
long-standing puzzle about the phase behavior of fluids adsorbed
in pores is now better understood.\\

{\underline{\bf{Acknowledgments}}}: We are grateful to the
Deutsche Forschungsgemeinschaft (DFG) for partial support (grant
N$^o$ TR6/A5 and C4) and thank the NIC J\"ulich for a generous grant of
computer time.


%
%

%




\begin{thebibliography}{99}
\bibitem{1} L.D. Gelb, K.E. Gubbins, R. Radhakrishnan, and M.
Sliwinska-Bartkowiak, Rep. Progr. Phys. {\bf 62}, 1573 (1999)

\bibitem{2} M. Sch\"on and S. Klapp, {\it Reviews in Computational
Chemistry, Vo. 24} (Wiley-VCH, Hoboken, 2007)

\bibitem{3} I. Brovchenko and A. Oleinikova, {\it Interfacial and
Confined Water} (Elsevier, Amsterdam, 2008)

\bibitem{4} S.J. Gregg and K.S.W. Sing, {\it Adsorption, Surface
Area, and Porosity} (Academic Press, New York, 2$^{\rm nd}$ ed.,
1982)

\bibitem{5} A.J. Liapis (ed.) {\it Fundamentals of Adsoprtion}
(Engineering Foundation, New York, 1987)


\bibitem{6} J. Fraissard (ed.) {\it Physical Adsorption, Theory,
and Applications} (Kluwer Acad. Publ., Dordrecht, 1997)

\bibitem{7} F. Rouquerol, J. Rouquerol, and K.S.W. Sing, {\it
Adsorption by Powders and Porous Solids: Principles, Methodology,
and Applications} (Academic Press, San Diego, 1999)

\bibitem{8} T. Thorsen, S.J. Maerkl, and S.R. Quake, Science
{\bf298}, 580 (2002)

\bibitem{9} A. Meller, J. Phys.: Condens. Matter {\bf 15}, 581
(2003)

\bibitem{10} E.I. Wolf, {\it Nanophysics and Nanotechnology}
(Wiley-VCH, Weinheim, Germany, 2004)

\bibitem{11} I.M. Squires and S.R. Quake, Rev. Mod. Phys. {\bf77},
977 (2005)

\bibitem{12} J.S. Rowlinson and B. Widom, {\it Molecular Theory of
Capillarity} (Oxford Univ. Press, Oxford, 1982)

\bibitem{13} C.A. Croxton (ed.) {\it Fluid Interfacial Phenomena}
(Wiley, New York, 1985)

\bibitem{14} I. Charvolin, J.-F. Joanny, and J. Zinn-Justin (eds.)
{\it Liquids at Interfaces} (North-Holland, Amsterdam, 1990)

\bibitem{15} D. Henderson (ed.) {\it Fundamentals of Inhomogeneous
Fluids} (M. Dekker, New York, 1992)

\bibitem{16} K. Binder, D.P. Landau, and M. M\"uller, J. Stat. Phys.
{\bf 110}, 1411 (2003)

\bibitem{17} P. Wiltzius, S.B. Dierker, and B.S. Dennis, Phys.
Rev. Lett. {\bf 62}, 804 (1989)

\bibitem{18} M.Y. Lin, S.K. Sinha, J.M. Drake, X.-I. Wu, P.
Thiyagarajan, and H.B. Stanley, Phys. Rev. Lett. {\bf 72}, 2207
(1994)

\bibitem{19} E. Kierlik, P.A. Monson, M.I. Rosinberg, and G.
Tarjus, J. Phys.: Condens. Matter {\bf14}, 9295 (2002)

\bibitem{20} S. Inoue, N. Ichikuni, T. Suzuki, T. Uematsu, and K.
Kaneko, J. Phys. Chem. B {\bf 102}, 4689 (1998)

\bibitem{21} M. Meyyappan (ed.) {\it Carbon Nanotubes: Science and
Applications} (CRC Press, Boca Raton, 2004)

\bibitem{22} Z.N. Yu, H. Gao, W. Wu, H.X. Ge, and S.Y. Chou, J.
Vac. Sci. Technol. B {\bf21}, 2874 (2003)

\bibitem{23} W. Reisner, K.J. Morton, R. R\"uhn, Y.M. Wang, Z. Yu,
M. Rosen, J.C. Sturm, S.Y. Chou, E. Frey, and R.H. Austin, Phys.
Rev. Lett. {\bf 94}, 196101 (2005)

\bibitem{24} W.T. Thomson (Lord Kelvin), Philos. Mag. {\bf42}, 448
(1871)

\bibitem{25} M.E. Fisher and H. Nakanishi, J. Chem. Phys. {\bf
75}, 5857 (1981)

\bibitem{26} H. Nakanishi and M.E. Fisher, J. Chem. Phys. {\bf
78}, 3279 (1983)

\bibitem{27} R. Evans and P. Taranzona, Phys. Rev. Lett. {\bf52},
557 (1984)

\bibitem{28} R. Evans, U. Marini Bettolo Marconi, and P.
Taranzona, J. Chem. Soc. Faraday Trans. {\bf2}, 1763 (1986)

\bibitem{29} G. Heffelfinger, F. Swol, and K. Gubbins, Mol. Phys.
{\bf 60}, 1381 (1987)

\bibitem{30} R. Evans, J. Phys.: Condens. Matter {\bf46}, 9899
(1990)

\bibitem{31} H. Dominguez, M.P. Allen, and R. Evans, Mol. Phys.
{\bf 96}, 209 (1999)

\bibitem{32} S. Varga, D. Boda, D. Henderson, and S. Sokolowski,
J. Colloid Interface Sci. {\bf227}, 223 (2000)

\bibitem{33} I. Brovchenko, A. Geiger, and D. Paschek, Fluid Phase
Equil. {\bf183}, 331 (2001)

\bibitem{34} M. Schmidt, A. Fortini, and M. Dijkstra, J. Phys.:
Condens. Matter {\bf16}, 4159 (2004)

\bibitem{35} B. Lefevre, A. Saugey, J.L. Barrat, L. Bocquet, E.
Charlaix, P.F. Gobin and G. Vigier, J. Chem. Phys. {\bf 120}, 4927
(2004)

\bibitem{36} N. Desbiens, I. Demachy, A. Fuchs, H.
Kirsch-Rodeschini, M. Soulard and J. Patarin, Angew. Chem. {\bf
117}, 5444 (2005)

\bibitem{37} P.G. de Gennes, Rev. Mod. Phys. {\bf 57}, 825 (1985)

\bibitem{38} D.E. Sullivan and M.M. Telo da Gama, in {\it Fluid
Interfacial Phenomena} (C.A. Croxton, ed.) p. 45 (Wiley, New York,
1986)

\bibitem{39} S. Dietrich, in {\it Phase Transitions and Critical
Phenomena, Vol. XII} (C. Domb and J.L. Lebowitz, eds.) p. 1
(Academic, New York, 1988)

\bibitem{40} M. Schick, in {\it Liquids at Interfaces} (J.
Charvolin, J.F. Joanny and J. Zinn-Justin, eds.) p. 415
(North-Holland, Amsterdam, 1990)

\bibitem{41} D. Chatain, Ann. Rev. Mat. Res. {\bf38}, 45 (2008)

\bibitem{42} D. Qu\'er\'e, Ann. Rev. Mat. Res. {\bf38}, 71 (2008)


\bibitem{43} S. Herminghaus, M. Brinkmann, and R. Seemann, Ann.
Rev. Mat. Res. {\bf38}, 101 (2008)

\bibitem{44} K. Binder, Ann. Rev. Mat. Res. {\bf38}, 123 (2008)

\bibitem{45} D.H. Everett and J.M. Haynes, J. Colloid Interface
Sci. {\bf38}, 125 (1972)

\bibitem{46} W.F. Saam and M.W. Cole, Phys. Rev. B {\bf11}, 1086
(1975)

\bibitem{47} B.V. Derjaguin and N.V. Churaev, J. Colloid Interface
Sci. {\bf54}, 157 (1976)

\bibitem{48} G. Mason, Proc. R. Soc. London, Ser. A {\bf390}, 47
(1983)

\bibitem{49} G.S. Heffelfinger, F. van Swol, and K.E. Gubbins, J.
Chem. Phys. {\bf89}, 5202 (1988)

\bibitem{50} P.C. Ball and R. Evans, Langmuir {\bf5}, 714 (1989)

\bibitem{51} C.G.V. Burgess, D.H. Everett and S. Nutall, Pure
Appl. Chem. {\bf 61}, 1845 (1989)

\bibitem{52} W.D. Machin, Langmuir {\bf10}, 1235 (1994)

\bibitem{53} M. Thommes and G.H. Findenegg, Langmuir {\bf10}, 4270
(1994); T. Michalski, A. Benini, and G.H. Findenegg, Langmuir
{\bf7}, 185 (1991)

\bibitem{54} M. Thommes, G.H. Findenegg, and M. Schoen, Langmuir
{\bf11}, 2137 (1995)

\bibitem{55} K. Morishige, H. Fujii, M. Uga, and D. Kinukawa,
Langmuir {\bf13}, 3494 (1997)

\bibitem{56} K. Morishige and M. Shikimi, J. Chem. Phys. {\bf108},
7821 (1998)

\bibitem{57} A.V. Neimark, P.I. Ravikovich, and A. Vishnyakov,
Phys. Rev. E {\bf62}, R1493 (2000)

\bibitem{58} A. Vishnyakov and A.V. Neimark, J. Phys. Chem.
B {\bf105}, 7009 (2001)

\bibitem{59} K.G. Kornev, I.K. Shingareva, and A.V. Neimark,
Ad. Coll. Interface Sci. {\bf96}, 143 (2002)

\bibitem{60} K. Morishige and M. Ito, J. Chem. Phys. {\bf117},
8036 (2002)

\bibitem{61} L.D. Landau and E.M. Lifshitz, {\it Statistical
Physics, 3$^{rd}$ ed.} (Pergamon Press, Oxford, 1959)

\bibitem{62} V. Privman and M.E. Fisher, J. Stat. Phys. {\bf33},
385 (1983)

\bibitem{63} M.N. Barber, in {\it Phase Transitions and Critical
Phenomena} edited by C. Domb and J.L. Lebowitz (Academic, London,
1983) Vol. 8, Ch. 2

\bibitem{64} L.D. Gelb and K.E. Gubbins, Phys. Rev. E {\bf56}, 3185
(1997)

\bibitem{65} W.D. Machin, Langmuir {\bf15}, 169 (1999)

\bibitem{66} G.S. Heffelfinger, Z. Tan, K.E. Gubbins, U. Marini
Bettolo Marconi, and F. van Swol, Mol. Simul. {\bf2}, 393 (1989)

\bibitem{67} A.J. Liu, D.J. Durian, E. Herbolzheimer, and S.A.
Safran, Phys. Rev. Lett. {\bf65}, 1897 (1990)

\bibitem{68} I. Brovchenko, A. Geiger, and A. Oleinikova, Phys.
Chem. Chem. Phys. {\bf3}, 1567 (2001)

\bibitem{69} I. Brovchenko, A. Geiger, and A. Oleinikova, J.
Phys.: Condens. Matter {\bf16}, S5345 (2004)

\bibitem{70} S. Asakura and Oosawa, J. Chem. Phys. {\bf22}, 1255
(1954)

\bibitem{71} M.E. Fisher, J. Phys. Soc. Jpn. Suppl. {\bf26}, 87
(1969)

\bibitem{72} A.E. Ferdinand and M.E. Fisher, Phys. Rev. {\bf 185},
832 (1969)

\bibitem{73} B.M. McCoy and T.T. Wu, {\it The two-dimensional
Ising model} (Harvard University Press, Cambridge, Mass., 1973)

\bibitem{74} H. Au-Yang, and M.E. Fisher, Phys. Rev. B {\bf11},
3469 (1975)

\bibitem{75} H. Au-Yang and M.E. Fisher, Phys. Rev. B {\bf21}, 3956
(1980)

\bibitem{76} M.E. Fisher and H. Au-Yang, Physica A {\bf101}, 255
(1980)

\bibitem{77} G.C. Cabrera, R. Jullien, E. Br\'ezin and J.
Zinn-Justin, J. Physique {\bf47}, 1305 (1986)

\bibitem{78} E.V. Albano, K. Binder, D.W. Heermann, and W. Paul,
Z. Phys. B {\bf 77}, 445 (1989)

\bibitem{79} E.V. Albano, K. Binder, D.W. Heermann, K. Binder, J.
Chem. Phys. {\bf91}, 3700 (1989)

\bibitem{80} E.V. Albano, K. Binder, D.W. Heermann, and W. Paul,
Surface Sci. {\bf 223}, 151 (1989)

\bibitem{81}V. Privman, in {\it Finite Size Scaling and Numerical
Simulation of Statistical Systems} (ed. by V. Privman) p.1 (World
Scientific, Singapore, 1990)

\bibitem{82} A.O. Parry and R. Evans, Physica A {\bf181}, 250 (1992)

\bibitem{83} J. Stecki, A. Maciolek and K. Olaussen, Phys. Rev.
B {\bf49}, 1092 (1993)

\bibitem{84} R. Evans and J. Stecki, Phys. Rev. B {\bf49}, 8842
(1994)

\bibitem{85} T.W. Burkhardt and E. Eisenriegler, Phys. Rev. Lett.
{\bf74}, 3189 (1995)

\bibitem{86} A. Maciolek and J. Stecki, Phys. Rev. B {\bf49}, 8842
(1996)

\bibitem{87}  E. Carlon, A. Drzewinski and J. Rogiers, Phys. Rev.
B {\bf58}, 5070 (1998)

\bibitem{88} P. Nowakowski and M. Napiorkowski, J. Phys. A: Math.
Theor. {\bf42}, 475005 (2009)

\bibitem{89} D.P. Landau and K. Binder, {\it A Guide to Monte
Carlo Simulation in Statistical Physics}, 3$^{\rm rd}$ ed.
(Cambridge Univ. Press, Cambridge, 2009)

\bibitem{90} K. Kawasaki, in {\it Phase Transitions and Critical
Phenomena, Vo. 2}, edited by C. Domb and M.S. Green (Academic,
London, 1972), Chap. 11.

\bibitem{91} L. Onsager, Phys. Rev. {\bf65}, 117 (1944)

\bibitem{92} U. Wolff, Phys. Rev. Lett. {\bf62}, 361 (1989)

\bibitem{93} L.W. Lee and A.P. Young, Phys. Rev. Lett. {\bf90},
227203 (2003)

\bibitem{94} M.E. Fisher, J. Stat. Phys. {\bf34}, 667 (1984)

\bibitem{95} K. Binder, Z. Phys. B {\bf43}, 119 (1981)

\bibitem{96} K. Binder, Eur. Phys. J. B {\bf64}, 307 (2008)

\bibitem{97} C.N. Yang, Phys. Rev. {\bf85}, 808 (1952)

\bibitem{98} K. Binder, Phys. Rev. A {\bf25}, 1699 (1982)

\bibitem{99} K. Binder, Rep. Progr. Phys. {\bf50}, 783 (1987)

\bibitem{100} K. Binder and P.C. Hohenberg, Phys. Rev. B {\bf6},
3461 (1972)

\bibitem{101} K. Binder and P.C. Hohenberg, Phys. Rev. B {\bf9},
2194 (1974)

\bibitem{102} A.M. Ferrenberg and D.P. Landau, Phys. Rev.
B {\bf44}, 5081 (1991)

\bibitem{103} K. Binder Physica {\bf62}, 508 (1972)

\bibitem{104} K. Binder, Thin Solid Films {\bf20}, 367 (1974)

\bibitem{105} D.P. Landau, Phys. Rev. B {\bf13}, 2997 (1976)

\bibitem{106} D.P. Landau, Phys. Rev. B {\bf14}, 255 (1976)

\bibitem{107} K. Binder and D.P. Landau, J. Chem. Phys. {\bf96},
1444 (1992)

\bibitem{108} D. Nicolaides and R. Evans, Phys. Rev. B {\bf39},
9336 (1989)

\bibitem{109} O. Dillmann, W. Janke, M. M\"uller and K. Binder, J.
Chem. Phys. {\bf 114}, 5853 (2001)

\bibitem{110} P. Virnau and M. M\"uller, J. Chem. Phys. {\bf120},
10925 (2004)

\bibitem{111} R.H. Swendsen and J.-S. Wang, Phys. Rev. Lett.
{\bf57}, 2607 (1986)

\bibitem{112} K. Binder and D.P. Landau, Phys. Rev. B {\bf30}, 1477
(1984)

\bibitem{113} C. Borgs and R. Kotecky, J. Stat. Phys. {\bf61}, 79
(1990)

\bibitem{114} A. J. Liu and G.S. Grest, Phys. Rev. A {\bf44}, 7894
(1991)

\bibitem{115} W.C. Poon and P.N. Pusey, {\it Observation,
Prediction, and Simulation of Phase Transitions in Complex
Fluids}, ed. M. Baus, I.F. Rull, and J.P. Ryckaert, (Kluwer Acad.
Publ., Dordrecht, 1995), pp. 3-51.

\bibitem{116} A.V. Blaaderen, Progr. Colloid Polym. Sci. {\bf
104}, 59 (1997)

\bibitem{117} A.K. Arora and B.V.R. Tata, Adv. Colloid Interface
Sci. {\bf78}, 49, (1998)

\bibitem{118} H. L\"owen, J. Phys.: Condens. Matter {\bf 13}, 415
(2001)

\bibitem{119} C. Likos, Phys. Repts. {\bf348}, 267 (2001)

\bibitem{120} H.N.V. Lekkerkerker, W. Poon, P. Pusey, A.
Stroobants, and P. Warren, Europhys. Lett. {\bf20}, 559 (1992)

\bibitem{121} W.C. Poon, J. Phys.: Condens. Matter {\bf14}, 859
(2002)

\bibitem{122} D.G.A.L. Aarts, J.H. van der Wiel, and H.N.W.
Lekkerkerker, J. Phys.: Condens. Matter {\bf15}, S245 (2003)

\bibitem{123} D.G.A.L. Aarts, M. Schmidt, and H.N.W. Lekkerkerker,
Science , {\bf304} 847 (2004)

\bibitem{124} Y. Hennequin, D.G.A.L. Aarts, J.O. Indekeu, H.N.W.
Lekkerkerker, and D. Bonn, Phys. Rev. Lett. {\bf100}, 178305
(2008)

\bibitem{125} A. Vrij, Pure Appl. Chem. {\bf48}, 471 (1976)

\bibitem{126} M. Schmidt, A. Fortini, and M. Dijkstra, J. Phys.:
Condens. Matter {\bf15}, S3411 (2003)

\bibitem{127} R.L.C. Vink and J. Horbach, J. Chem. Phys. {\bf121},
3253 (2004)

\bibitem{128} R.L.C. Vink, J. Horbach, and K. Binder, Phys. Rev.
E{\bf71}, 011401 (2005)

\bibitem{129} R.L.C. Vink, J. Horbach, and K. Binder, J. Chem.
Phys. {\bf122}, 134905 (2005)

\bibitem{130} R.L.C. Vink, A. DeVirgiliis, K. Binder, and J. Horbach, Phys.
Rev. E {\bf73}, 056118 (2006)

\bibitem{131} A. Fortini, M. Schmidt, and M. Dijkstra, Phys. Rev.
E {\bf73}, 051502 (2006)

\bibitem{132} K. Binder, J. Horbach, R.L.C. Vink, and A.
DeVirgiliis, Soft Matter {\bf4}, 1555 (2008)

\bibitem{133} F. Wang, D. P. Landau, Phys. Rev. Lett. {\bf86}, 2050 (2001)

\end{thebibliography}
\end{document}